\documentclass[sigconf, nonacm]{acmart}

\newcommand\paperavailabilityurl{https://github.com/TsinghuaDatabaseGroup/prepbench}
\newcommand\paperpagestyle{plain}

\usepackage{xspace}

\usepackage{pifont}
\usepackage{colortbl}
\usepackage{makecell}

\usepackage{algorithm}
\usepackage{algorithmic}

\usepackage{amsmath}
\usepackage{amsthm}

\usepackage{enumitem}
\usepackage{caption}
\usepackage{tabularx}
\usepackage{booktabs}
\usepackage{array}
\usepackage{arydshln} 

\usepackage{adjustbox}

\usepackage{subcaption}

\usepackage{siunitx}
\usepackage{multirow}

\usepackage{placeins}

\usepackage{xcolor}

\usepackage[most]{tcolorbox}
\usepackage{listings}

\usepackage{wasysym}
\usepackage{tikz}

\newcommand{\figref}[1]{Figure~\ref{#1}}
\newcommand{\tabref}[1]{Table~\ref{#1}}
\newcommand{\secref}[1]{Section~\ref{#1}}

\newcommand{\system}[1]{\textsc{#1}}          

\newcommand{\sysname}{\system{PrepBench}\xspace}

\newcommand{\oreq}{\textit{original request}\xspace}
\newcommand{\Oreq}{\textit{Original request}\xspace}

\newcommand{\intab}{\textit{input tables}\xspace}

\newcommand{\gtout}{\textit{ground-truth output}\xspace}

\newcommand{\gtcode}{\textit{ground-truth code}\xspace}

\newcommand{\dreq}{\textit{disamb request}\xspace}

\newcommand{\dkb}{\textit{disamb KB}\xspace}

\newcommand{\gtflow}{\textit{ground-truth workflow}\xspace}

\newcommand{\orig}{Orig\xspace}
\newcommand{\disamb}{Disamb\xspace}
\newcommand{\interact}{Interact\xspace}
\newcommand{\profile}{Profile\xspace}

\newcommand{\todo}[1]{\textcolor{red}{\textbf{[TODO: #1]}}}

\definecolor{cmarkgreen}{RGB}{34,139,34}    
\definecolor{xmarkred}{RGB}{220,20,60}      
\definecolor{pmarkorange}{RGB}{255,140,0}   
\definecolor{headerblue}{RGB}{230,240,250}  
\newcommand{\cmark}{\textcolor{cmarkgreen}{\ding{51}}}  
\newcommand{\xmark}{\textcolor{xmarkred}{\ding{55}}}    

\definecolor{pbBlue}{RGB}{85,120,255}
\definecolor{pbGray}{RGB}{120,120,120}

\lstdefinestyle{prepcode}{
  basicstyle=\ttfamily\footnotesize,
  columns=fullflexible,
  breaklines=true,
  showstringspaces=false
}

\tcbset{
  pbCodeBox/.style={
    colback=white,
    colframe=pbBlue,
    boxrule=0.6pt,
    arc=0pt,
    left=6pt,right=6pt,top=4pt,bottom=4pt,
    boxsep=0pt,
    width=\columnwidth
  }
}

\definecolor{pbLightGray}{RGB}{242,242,242}
\newtcolorbox{takeawaybox}{
  enhanced,
  colback=pbLightGray,
  colframe=black!60,
  boxrule=0.6pt,
  sharp corners,
  width=\columnwidth,
  boxsep=0pt,
  left=5pt,
  right=5pt,
  top=4pt,
  bottom=4pt,
  before skip=5pt,
  after skip=5pt
}

\begin{document}

\makeatletter
\g@addto@macro\UrlSpecials{\do\,{\mathchar`\,\penalty\UrlBreakPenalty\hbox{ }}}
\makeatother

\title{PrepBench: How Far Are We from Natural-Language-Driven\texorpdfstring{\\}{ }Data Preparation?}

\author{Jingzhe Xu, Rui Wang, Jiannan Wang, Guoliang Li}
\affiliation{%
  \institution{Department of Computer Science and Technology, BNRist, Tsinghua University}
  \city{Beijing}
  \country{China}
}
\email{xjz@bit.edu.cn, akane@hust.edu.cn, jnwang@tsinghua.edu.cn, liguoliang@tsinghua.edu.cn}

\begin{abstract}

Data preparation is a central and time-consuming stage in data analysis workflows.
Traditionally, commercial tools have relied on graphical user interfaces (GUIs) to simplify data preparation, allowing users to define transformations through visual operators and workflows.
Recent advances in large language models (LLMs) raise the possibility of a paradigm shift toward natural language (NL)-driven data preparation, in which users can specify preparation intents in NL directly. However, it remains unclear how far current LLM-based agents are from this paradigm shift in practice.
Existing code generation benchmarks do not capture key characteristics of data preparation, including ambiguous user intents, imperfect real-world data, and the need to translate code into interpretable workflows for validation. To bridge this gap, we present PrepBench, a benchmark designed to evaluate NL-driven data preparation along three core capabilities: interactive disambiguation, prep-code generation, and code-to-workflow translation.
We crawl data from the Preppin\textquotesingle{} Data Challenges, and then extend it into a systematically designed benchmark. The benchmark covers diverse domains, and each task involves 3 to 18 data preparation steps. Nearly half of the tasks require over 100 lines of Python code, and the longest solutions approach 300 lines. 
Our evaluation shows that, despite recent progress, realizing this paradigm shift remains challenging for state-of-the-art LLMs. PrepBench provides a principled benchmark for measuring this gap and helps identify key challenges toward realizing NL-driven data preparation.

\end{abstract}

\maketitle

\pagestyle{\paperpagestyle}

\ifdefempty{\paperavailabilityurl}{}{
\begingroup\small\noindent\raggedright\textbf{Artifact Availability:}\\
The source code, data, and/or other artifacts have been made available at \url{\paperavailabilityurl}.
\endgroup
}


\begin{figure*}[t]
  \centering
  \includegraphics[width=\textwidth]{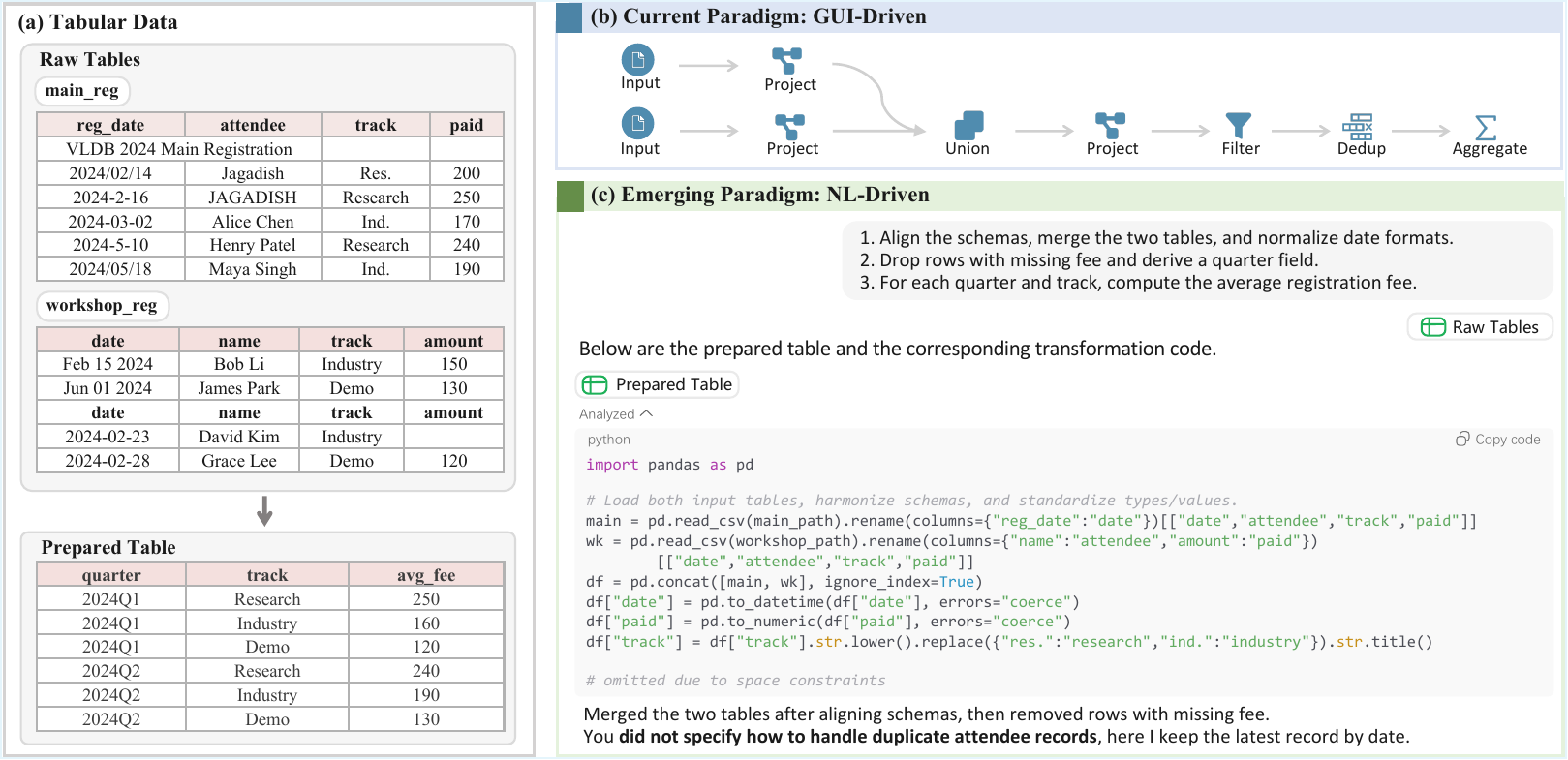}
  \Description{Comparison of GUI-driven and NL-driven data preparation paradigms.} 
  \caption{(a) Unprepared and prepared tables: Raw data from two tables alongside the resulting prepared data;
(b) GUI-driven data preparation: Users define a preparation workflow through visual operators (e.g., Project, Union, Filter); (c) NL-driven data preparation: Users specify tasks in natural language, and the system generates the corresponding code.}
  \label{fig:intro}
\end{figure*}

\section{Introduction}
\label{sec:intro}


Data preparation has long been considered a major bottleneck in data analysis~\cite{raman2001potter,kandel2011wrangler}. In practice, analysts often spend considerable time cleaning, transforming, and reshaping data before any downstream analysis can begin~\cite{anaconda2020state}.
To address this issue, the industry has adopted commercial data preparation tools such as Tableau Prep~\cite{tableau_prep_about} and SAS Data Preparation~\cite{sas_data_prep}. 
These tools allow users to construct data preparation workflows through graphical user interfaces (GUIs) by chaining visual operators (e.g., filter, join, aggregate, pivot).  
We refer to this paradigm as \emph{GUI-driven data preparation}. 
Figure~\ref{fig:intro}(a) shows an example in which two raw tables need to be transformed into a prepared form. Figure~\ref{fig:intro}(b) shows one workflow constructed by analysts to perform this data preparation.

While effective, this paradigm requires users to translate natural language intents into GUI operations, which can be indirect and error-prone. 
Moreover, GUI-based tools still impose a non-trivial learning curve despite being more accessible than programming.
Recent advances in large language models (LLMs) offer a possible way to close this gap~\cite{narayan2022can,fan2024autoprep}. 
As shown in Figure~\ref{fig:intro}(c), users can upload data to an LLM-based agent  (e.g., ChatGPT~\cite{chatgpt}) and directly express their data preparation intent in natural language (e.g., "Align the schemas, merge the two tables, and normalize date formats"). 
ChatGPT then generates executable code and produces the prepared data. 
We refer to this interaction model as \emph{NL-driven data preparation}. This paradigm can significantly lower the barrier to data preparation and make data analysis more accessible to a broader range of users.

As a result, one natural question is: \emph{how far are we from NL-driven data preparation?} To answer this question, we introduce \sysname, a new benchmark for NL-driven data preparation. Our goal is not only to conduct an end-to-end evaluation, but also to understand the fundamental capabilities that LLM-based systems must support to make NL-driven data preparation practical. Specifically, \sysname evaluates systems along three capabilities: (1) \emph{Interactive Disambiguation}, which tests whether systems can detect and resolve ambiguous intents through interaction; (2) \emph{Prep-Code Generation}, which tests whether systems can generate correct preparation code from user requests; and (3) \emph{Code-to-Workflow}, which tests whether generated code can be mapped to GUI workflows.


These capabilities are motivated by practical requirements, but have not been evaluated together in existing benchmarks. Prior benchmarks have studied NL interfaces for various data-related tasks, including NL-to-SQL~\cite{huo2025bird,dong2025practiq}, NL-to-Code~\cite{yin2023natural,lai2023ds}, NL-to-Pipeline~\cite{ge2025text,jin2025elt}, and Code-to-Workflow understanding~\cite{li2025crabs}. However, as summarized in Table~\ref{tab:bench-coverage}, these benchmarks either address different problem settings or focus on individual capabilities in isolation (see Section~\ref{sec:task:bench_compare}). To the best of our knowledge, \sysname is the first to systematically evaluate all three capabilities for NL-driven data preparation.

Evaluating these capabilities requires a benchmark based on realistic tasks. \sysname is constructed by crawling data from the Preppin\textquotesingle{} Data Challenges~\cite{preppindata_challenges}, a weekly series that provides practical data preparation tasks to Tableau Prep users. We then extend the collected data into a systematically designed benchmark.  \sysname consists of  306 tasks over 829 input tables and covers 32 application domains, including retail, sports, education, finance, and transportation.  The tasks are diverse and non-trivial: each task involves 3 to 18 data preparation steps, nearly half require more than 100 lines of Python code, and the longest solutions approach 300 lines.  Although derived from Tableau Prep training materials, the operations and workflows in \sysname are generic and supported by a wide range of commercial data preparation systems~\cite{hameed2020data}. These characteristics make \sysname well-suited for evaluating realistic tabular data preparation tasks.

It is worth noting that Preppin\textquotesingle{} Data alone cannot be directly used as a benchmark to evaluate our three target capabilities. There are two challenges that need to be addressed. 
First, Preppin\textquotesingle{} Data lacks executable ground-truth code and unambiguous task descriptions—both essential but expensive to create manually at scale. We address this by designing an agent-based pipeline that automatically generates ground-truth code and derives unambiguous task descriptions from it. This approach enables scalable benchmark construction with reduced manual effort.

The second challenge lies in evaluating interactive disambiguation. In real-world data preparation scenarios, NL requests are often ambiguous, requiring systems to identify missing information, ask clarification questions, and refine solutions based on user feedback. As part of benchmark construction, we build a disambiguation knowledge base that records ambiguity cases in the original requests together with their corresponding resolutions. We further design a \emph{user simulator} that automatically responds to clarification questions using this knowledge base. This setup simulates realistic user interaction, where a system detects ambiguity, requests clarification, and receives additional input to complete the task.

We conduct comprehensive experiments on \sysname across ten state-of-the-art proprietary and open-weight models, evaluating both end-to-end performance and each core capability. 
The results reveal several important insights. 
First, although current LLMs show strong code generation ability, the best model (GPT-5.1-Codex) achieves only 54.9\% accuracy on prep-code generation. 
Ambiguity has a large impact: removing it improves accuracy to 85.3\%. 
Second, interactive disambiguation helps, but current LLMs frequently ask incomplete or ineffective clarification questions, which limits its benefits. 
Third, code-to-workflow translation remains a major bottleneck: even when correct prep-code is generated, the resulting workflows often do not match the intended data preparation steps.
Finally, higher model cost does not consistently predict better performance: a lightweight model (Gemini~3~Flash) nearly matches the best model's accuracy at less than one-fifth of the cost. 
Overall, these results indicate that NL-driven data preparation is promising but remains challenging in practice.
They also demonstrate the importance of \sysname as a benchmark for systematically revealing limitations that are not captured by existing evaluations.

In summary, the paper makes the following contributions:

\begin{itemize}[leftmargin=.8em]
    \item We present \sysname, a new benchmark that systematically evaluates NL-driven data preparation across three core capabilities: interactive disambiguation, prep-code generation, and code-to-workflow translation.
    \item We construct \sysname by extending the Preppin\textquotesingle{} Data Challenges with executable ground-truth data, disambiguation knowledge base, and unambiguous task specifications. 
    \item We design three execution modes that enable both end-to-end evaluation and fine-grained analysis of individual capabilities.
    \item We conduct a comprehensive evaluation of LLM-based agents on \sysname, identifying key limitations and highlighting opportunities for future research.
\end{itemize}

Section~\ref{sec:task} defines NL-driven data preparation. Sections~\ref{sec:benchmark} and~\ref{sec:capability} then detail the construction of \sysname and its corresponding execution modes. Our experimental findings and subsequent research opportunities are discussed in Sections~\ref{sec:experiments} and~\ref{sec:research-opportunity}, respectively. Finally, we review related work in Section~\ref{sec:related} and conclude in Section~\ref{sec:conclusion}.

\section{NL-Driven Data Preparation}
\label{sec:task}

We first define NL-driven data preparation (\secref{sec:task:formulation}), then present three required capabilities (\secref{sec:task:capabilities}), and finally position \sysname against existing benchmarks (\secref{sec:task:bench_compare}).

\subsection{Problem Statement}
\label{sec:task:formulation}

Given a set of raw tables $D$ and an NL request $q$, the goal of NL-driven data preparation is to transform $D$ into one or multiple output tables according to the specifications in $q$.
In this paper, \emph{data preparation} refers to structural and syntactic table transformations,
such as joining, reshaping, filtering, and formatting (e.g., parsing dates and normalizing strings).
We do not target semantic data repair that cannot be determined from $D$ and $q$ alone, 
such as inferring missing values and correcting out-of-date values.

To illustrate, Figure~\ref{fig:intro}(c) depicts a typical NL-driven data preparation scenario. The input $D$ comprises two raw tables, \texttt{main\_reg} and \texttt{workshop\_reg}, which record registration data for a main conference and its workshops. These tables exhibit structural heterogeneity (e.g., schema mismatches like attendee vs. name) and syntactic inconsistencies (e.g., mixed date formats). The user provides an NL request $q$—"Align the schemas... Drop rows... compute the average registration fee". Guided by $q$, the system transforms $D$ into the target output by generating executable code that unifies schemas and standardizes values (e.g., normalizing "Ind." to "Industry").

\subsection{Three Fundamental Capabilities}
\label{sec:task:capabilities}

Accurate code generation alone does not make NL-driven data preparation practical. This is because:
(1) user requests are often ambiguous and can be interpreted in multiple ways;
(2) real datasets are often too large to examine thoroughly, and the code written for the sample may miss important data issues; and
(3) the generated code is hard for users, who are familiar with GUI-based workflows, to understand and check.
Therefore, a practical NL-driven data preparation system requires the following three capabilities.


\noindent\textbf{Interactive Disambiguation.} Consider the example in Figure~\ref{fig:intro}(c). The NL request does not specify how to handle duplicate attendees in the \texttt{main\_reg} table (e.g., "keep the latest record" vs. "keep the one with the highest fee"). A wrong guess here may introduce silent errors into the data. The interactive disambiguation capability addresses this by detecting such ambiguities and resolving them through user clarification instead of default assumptions.

\noindent\textbf{Prep-Code Generation.} After disambiguation, the system generates executable prep-code. In the above example, a user might only see standard dates (e.g., ``2024/02/14'') in the sample of the \texttt{main\_reg} table, but row 10,000 might contain a different format (``February 15 2024''). The prep-code generated solely based on the sample will fail. The prep-code generation capability addresses this by profiling the full dataset to discover hidden irregularities and generating robust code that handles all variations.


\noindent\textbf{Code-to-Workflow.}
After generating prep-code and prepared tables, users often need to verify the transformation logic. However, verifying raw code can be challenging; for instance, confirming that a complex Pandas script correctly dropped ``missing fee'' rows may be difficult for non-experts. In contrast, validating a visual ``Filter'' node in a workflow diagram is typically more intuitive. 
The \textit{workflow translation} capability addresses this by translating generated code into an interpretable workflow (as shown in Figure~\ref{fig:intro}(b)), which can be run on the input tables to facilitate verification.



\subsection{Comparison with Existing Benchmarks}
\label{sec:task:bench_compare}

\begin{table}[t]
  \begingroup

  \centering
  \footnotesize
  \setlength{\tabcolsep}{2pt}
  \renewcommand{\arraystretch}{1.08}
  \caption{Comparison with representative benchmarks (\cmark: evaluated; \xmark: not targeted; parentheses specify the setting).}
  \label{tab:bench-coverage}

  \resizebox{\columnwidth}{!}{%
  \begin{tabular}{@{}c|l|lll@{}}
    \hline
    \multicolumn{1}{c|}{\multirow[c]{2}{*}{\rule{0pt}{4.8ex}\textbf{Benchmark}}} &
    \multicolumn{1}{c|}{\multirow[c]{2}{*}{\rule{0pt}{4.8ex}\textbf{Task}}} &
    \multicolumn{3}{c}{\textbf{Capability}\rule{0pt}{2.1ex}} \\
    \cline{3-5}
    & &
    \makecell[l]{\rule{0pt}{2.4ex}\textbf{Interactive}\\\textbf{Disambig.}} &
    \makecell[l]{\rule{0pt}{2.4ex}\textbf{Prep-Code}\\\textbf{Generation}} &
    \makecell[l]{\rule{0pt}{2.4ex}\textbf{Workflow}\\\textbf{Translation}} \\
    \hline

    BIRD-INTERACT &
    NL2SQL &
    \makecell[l]{\cmark (SQL)} &
    \xmark &
    \xmark \\

    PRACTIQ &
    NL2SQL &
    \makecell[l]{\cmark (SQL)} &
    \xmark &
    \xmark \\

    DS-1000 &
    NL2Code &
    \xmark &
    \makecell[l]{\cmark (cell)} &
    \xmark \\

    ARCADE &
    NL2Code &
    \xmark &
    \makecell[l]{\cmark (cell)} &
    \xmark \\

    PARROT &
    NL2Pipeline &
    \xmark &
    \makecell[l]{\cmark (schema + cell)} &
    \xmark \\

    ELT-Bench &
    NL2Pipeline &
    \xmark &
    \makecell[l]{\cmark (schema)} &
    \xmark \\

    CRABS &
    Code2Workflow &
    \xmark &
    \xmark &
    \makecell[l]{\cmark (descriptive)} \\

    \hline

    \textbf{\sysname} &
    \makecell[l]{\textbf{NL2Code}\\\textbf{Code2Workflow}} &
    \makecell[l]{\cmark~(\textbf{prep-code})} &
    \makecell[l]{\cmark~(\textbf{schema + cell})} &
    \makecell[l]{\cmark~(\textbf{descriptive}\\\textbf{+ executable})} \\

    \hline
  \end{tabular}%
  }

  \arrayrulecolor{black}
  \endgroup
\end{table}

\tabref{tab:bench-coverage} compares \sysname with relevant benchmarks.

\noindent\textbf{Benchmarks for Interactive Disambiguation.} 
Recent benchmarks such as BIRD-INTERACT~\cite{huo2025bird}  and PRACTIQ~\cite{dong2025practiq} introduce interaction into NL-to-SQL tasks. 
BIRD-INTERACT and PRACTIQ both evaluate ambiguity handling in NL-to-SQL.
BIRD-INTERACT lets the system ask follow-up questions during evaluation, while PRACTIQ provides a fixed clarification dialogue for each ambiguous task.
Both benchmarks use clarification to recover the intended SQL query.
In contrast, \sysname evaluates disambiguation for data preparation, where clarification concerns how raw tables should be cleaned, transformed, and combined, and the resolved intent is used to generate prep-code and a workflow.

\noindent\textbf{Benchmarks for Prep-Code Generation.} 
Benchmarks such as DS-1000~\cite{lai2023ds} and ARCADE~\cite{yin2023natural} evaluate NL-to-code generation for data analysis. 
DS-1000 tasks are typically short snippets, with 3.6 lines on average, while ARCADE generates code for the next cell in a data science notebook.
These benchmarks may involve cell-level irregularities, but they do not evaluate schema-level irregularities which are common in real-world data preparation.
PARROT~\cite{ge2025text} generates executable pipelines rather than code snippets and covers both schema-level and cell-level irregularities, but it does not evaluate clarification for ambiguous requests or code-to-workflow translation.
ELT-Bench~\cite{jin2025elt} evaluates end-to-end ELT pipeline construction with schema-level irregularities, but does not target cell-level dirty values.
In contrast, \sysname evaluates prep-code generation for multi-step data preparation workflows with both schema-level and cell-level irregularities.

\noindent\textbf{Benchmarks for Code-to-Workflow.} 
CRABS~\cite{li2025crabs} extracts dependency graphs from notebook code to support interpretability, capturing information flow and cell execution dependencies.
These graphs provide descriptive workflow representations that help users reason about code.
\sysname builds on this insight but focuses on NL-driven data preparation.
In this setting, workflows are evaluated as both descriptive and executable artifacts, supporting inspection as well as validation against data.

\section{PrepBench}
\label{sec:benchmark}

This section presents \sysname. 
We first describe the data source (\secref{sec:benchmark:source}), and then discuss how to extend it into \sysname (\secref{sec:benchmark:construction}). 
We finally report benchmark statistics (\secref{sec:coverage_difficulty}).

\subsection{Data Source}
\label{sec:benchmark:source}

Preppin\textquotesingle{} Data~\cite{preppindata_challenges} is a collection of data preparation challenges based on tabular datasets. Each challenge requires applying multiple preparation steps to turn raw data into analysis-ready data. Since 2019, the collection has grown to over 300 challenges, covering a wide range of practical tasks.

Preppin\textquotesingle{} Data has three characteristics that make it well suited as the basis for our benchmark. First, it offers broad coverage: the challenges span 32 application domains, including retail, sports, education, finance, and transportation, capturing a wide range of data preparation scenarios. Second, the tasks are non-trivial. They typically require multi-step transformations over irregular data and are substantially more complex than small, isolated operations. Third, the tasks are realistic. They are derived from practical data preparation problems and reflect the skills required in real analytics workflows. As a result, performance on these tasks is a meaningful indicator of real-world data preparation capability.

\subsection{Benchmark Construction}
\label{sec:benchmark:construction}

We convert each Preppin\textquotesingle{} Data weekly challenge into a benchmark task through four stages. We (i) parse the challenge pages and clean the extracted task data and descriptions, (ii) generate executable code that reproduces the ground-truth output, (iii) use the verified code as a reference to disambiguate the request, and (iv) translate the code into a workflow.

\noindent\textbf{Task Parsing and Cleaning.}
Each challenge is published as a blog post containing a scenario, step-by-step instructions, and the expected output schema, along with downloadable input and output tables.
We parse each page to extract three assets: \oreq (scenario, steps, and output schema), \intab, and \gtout.
\Oreq is cleaned by removing non-task text such as announcements, navigation links, and author information.
We exclude one out-of-scope challenge (2024 Week 49), which is not a table preparation task.

\noindent\textbf{Ground-Truth Code Generation.} If we were only interested in evaluating end-to-end accuracy, generating ground-truth code would not be necessary. By using the ground-truth output, we could directly compare it to the system-generated output to determine if the data preparation code is correct. However, since we aim to evaluate the three core capabilities independently, as will be seen in later sections, ground-truth code is essential for each capability. 

We first consider two approaches to generate ground-truth code, but both have limitations. First, although Preppin\textquotesingle{} Data provides solution posts, they are expressed as Tableau Prep workflows stored in a proprietary format (.tflx) and depend on Tableau’s closed-source runtime. This makes it difficult to reliably translate them into ground-truth code. Second, manually writing ground-truth code is also challenging. These weekly challenges are often complex, and producing correct code is time-consuming. Moreover, NL requests are frequently ambiguous, requiring repeated interpretation, code revision, and output comparison to ensure correctness.

We adopt an agent-based pipeline to generate \gtcode. The agent performs two tasks: profiling tables to summarize their characteristics, and synthesizing code based on the request. The profiling process is described in \secref{sec:prep_code_generation}. 
Unlike a standard NL-driven data preparation task, here the ground-truth output is available.
This gives the agent direct feedback on whether the generated code is correct and allows it to revise the code automatically.

Specifically, our agent pipeline proceeds in two phases:

\begin{enumerate}[leftmargin=*]
\item \textbf{Initial Phase}: The agent begins by reading the original request, profiling the relevant tables, and generating the code. The generated code is then executed, and its output is compared to the ground-truth output.
\item \textbf{Iterative Phase}: If the generated output does not match the ground-truth output, the agent analyzes the differences (such as row mismatches, missing keys, or value differences). Using this feedback, the agent revises the code and executes it again. This process continues until the output matches the expected result or the iteration limit is reached.
\end{enumerate}


This method works for 57.3\% of tasks, showing that it can generate the right code most of the time. 
For the remaining tasks, the feedback helps narrow down the possible issues, and we complete the code through manual debugging.

\begin{figure}[t]
  \centering
  \includegraphics[width=\columnwidth]{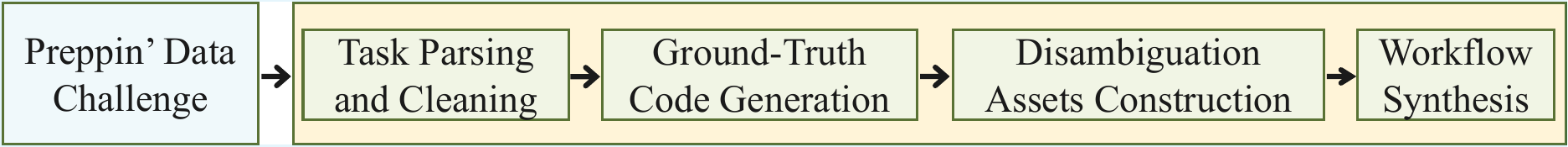}
 \caption{An overview of \sysname construction.}
  \Description{Flow diagram of the benchmark construction process from source materials to validated tasks and assets.}
  \label{fig:construction}
\end{figure}

\noindent\textbf{Disambiguation Assets Construction.} To evaluate interactive disambiguation, we need to figure out where an NL request can be interpreted in multiple reasonable ways and how such ambiguity can be resolved through interaction. To support this evaluation, we construct two disambiguation-related assets using the ground-truth code as a reference.
The first asset is a disambiguated request (abbreviated as \dreq). It rewrites the original request so that it has no ambiguities. The second asset is a disambiguation knowledge base (abbreviated as \dkb). It records ambiguity cases observed in the original request together with one corresponding resolution. In this section, we focus on ambiguities related to data transformations. Issues caused by irregular input data are handled separately in the profiling stage (\secref{sec:prep_code_generation}).

\noindent{\underline{\dreq}.} We construct \dreq through a validation-based process. We first prompt an LLM (GPT-5.1) to rewrite the \oreq  by clarifying ambiguous parts. This produces a candidate \dreq. We then assess whether the rewritten request is sufficiently clear for downstream use. Specifically, we ask a strong coding model (GPT-5.1-Codex) to generate code using only the rewritten request and the \intab, and compare the resulting output with the \gtout. If the outputs match, we treat the rewritten request as an acceptable \dreq.

If the outputs do not match, we manually inspect the failure and assign it to one of three cases. (i) The rewritten request allows multiple interpretations. (ii) The rewritten request is clear, but irregular input data leads to execution failures or output mismatches. (iii) The rewritten request is clear, but the coding model fails to produce correct code. 
Among these cases, only (i) indicates that the request may still contain ambiguity. In such cases, we revise the rewritten request and repeat the validation process until the remaining failures no longer fall into case (i).
For cases (ii) and (iii), the mismatch does not indicate ambiguity in the rewritten request, so we accept the \dreq as final without further revision.

\noindent{\underline{\dkb}.} We construct \dkb by carefully aligning ambiguous parts of the original request with their clarified counterparts in \dreq. Each entry in \dkb captures a concrete ambiguity case. It includes (a) the specific text in the original request that can be interpreted in multiple reasonable ways, (b) the corresponding clarified text in \dreq that reflects one intended interpretation, and (c) a minimal code snippet extracted from the ground-truth code that implements this interpretation. 

We manually review extracted entries to ensure that they correspond to valid ambiguity cases, and that the recorded resolutions are consistent with the ground-truth code. This manual review also helps ensure that different ambiguity cases are treated consistently across tasks. \secref{sec:interactive_disambiguation} further describes the ambiguity taxonomy, the structure of \dkb, and how these assets are used to evaluate interactive disambiguation.

\noindent\textbf{Workflow Synthesis.} To evaluate the code-to-workflow capability, we need a workflow representation that allows users to inspect and verify data preparation logic. We define a workflow abstraction with a fixed set of operators and an execution engine that applies a workflow to \intab and produces output tables. For each task, we construct a ground-truth workflow corresponding to \gtcode. We first prompt an LLM with the operator definitions and \gtcode to generate a \emph{candidate} workflow, and then validate it by executing the workflow and comparing its output with the ground-truth output. If the \emph{candidate} produces an incorrect result, we manually revise it until the outputs match. The final validated workflow is recorded as \gtflow.

Since every task admits a validated \gtflow, the operator set is sufficient to express the transformations required by \sysname. \secref{sec:code_to_gui_translation} describes the operators, execution engine, and evaluation protocol.

LLMs assist the construction of \gtcode, \dreq, \dkb, and \gtflow.
Although execution-based validation and manual review reduce construction errors and mitigate LLM-induced bias, they cannot eliminate all remaining risks.
(i) For \gtcode and \gtflow, execution is the main check.
We accept them only when their execution on \intab yields an output table that matches \gtout, which is provided by Preppin' Data.
The remaining risk is implementation preference, since the accepted code or workflow may reflect one LLM-preferred way to realize the transformation.
(ii) For \dreq, we check whether a coding model can reproduce \gtout from the rewritten request alone.
This check helps detect remaining ambiguity, but it does not prove that \dreq is fully disambiguated, since the model may guess underspecified choices.
When a rewrite remains ambiguous, we manually revise it.
(iii) For \dkb, entries link ambiguous spans in \oreq to clarifications in \dreq and relevant snippets in \gtcode.
This process may miss some ambiguity cases or align them imperfectly with their clarifications.
We manually review entries for validity and consistency with \gtcode.
For valid clarification questions not covered by \dkb, the simulator refers to \gtcode so the agent can still receive an appropriate response.
A remaining risk is that LLM-assisted artifacts may reflect the style of the construction model, which could give models from a similar family an advantage in evaluation.

\begin{figure}[t]
  \centering
  \includegraphics[width=\columnwidth]{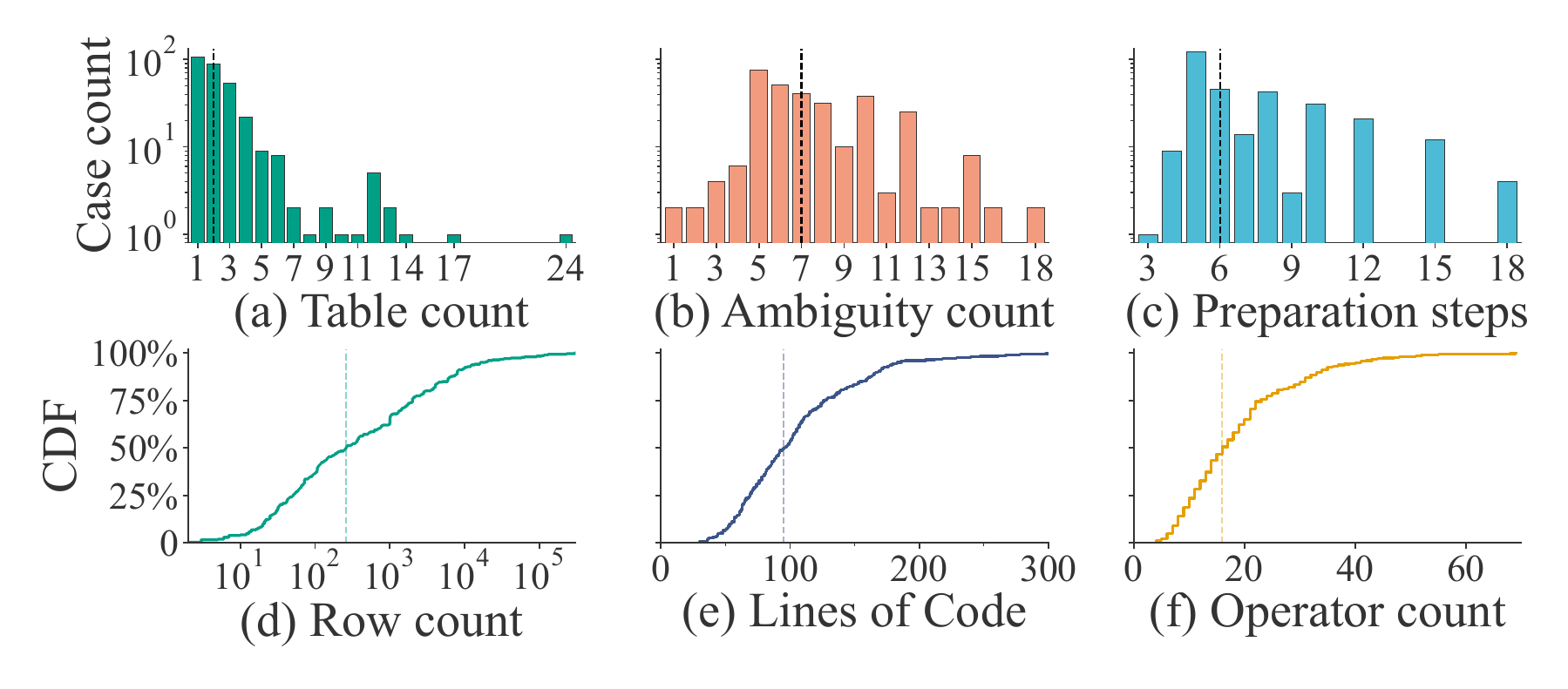}
 \caption{\sysname statistics.}
  \Description{Summary charts reporting task domains, preparation steps, ambiguity distribution, and data irregularity statistics in the benchmark.}
  \label{fig:statistic}
  \vspace{-1em}
\end{figure}

\subsection{Benchmark Statistics}
\label{sec:coverage_difficulty}

\sysname comprises 306 realistic tasks over 829 input tables across 32 domains. We report detailed benchmark statistics to better characterize the benchmark.

\noindent\textbf{Input-Data Scale.}
To explore the input data scale, we count \intab and their total rows per task.
As shown in \figref{fig:statistic}(a), most tasks involve at most six tables, while some require integrating many tables (up to 24).
\figref{fig:statistic}(d) further shows substantial diversity in input size. 
Over 30\% of tasks exceed 1{,}000 rows, and the largest inputs reach 291{,}287 rows.
These results show that \sysname\ demands multi-table processing and scalable data handling.

\noindent\textbf{User-Request Diversity.} We characterize the diversity of user requests by the distribution of ambiguities in \oreq and preparation steps in \dreq. As illustrated in \figref{fig:statistic}(b) and (c), \sysname covers a broad spectrum of difficulty. The tasks range from simple requests with few ambiguities and steps (e.g., $\leq3$) to highly complex ones exceeding 15 ambiguities or 12 steps. These statistics indicate that \sysname evaluates systems on a diverse mix of scenarios.

\noindent\textbf{Preparation-Step Coverage.}
Finally, we map each task's ground-truth implementation to a fixed set of preparation step types (Table~\ref{tab:feature_coverage}) and report which types are covered in \sysname.
\sysname provides broad coverage of preparation step types, including core steps such as \emph{Derive column}, \emph{Change column data type}, and \emph{Join \& union}, reflecting common needs like computed fields, format normalization, and multi-table integration.
It also includes less frequent operations such as \emph{Pivot \& unpivot}, \emph{Split column}, and \emph{Merge columns} for reshaping and column-level restructuring.
We exclude several step types, such as those that rely on external data or operate on file encodings rather than the input tables.
Overall, \sysname emphasizes robustness to diverse step types and their multi-step composition.


\noindent\textbf{Solution Complexity.}
To assess solution complexity, we count the number of lines in \gtcode and the number of operators in \gtflow.
As shown in \figref{fig:statistic}(e), nearly half of the tasks require more than 100 lines of \gtcode, and the longest solutions approach 300 lines.
\figref{fig:statistic}(f) shows a similar distribution for workflows, with a median of about 20 operators and a maximum of 69.
These statistics indicate that many \sysname\ tasks require multi-step data-preparation solutions.





{
\renewcommand{\cmark}{\ding{51}}
\renewcommand{\xmark}{\ding{55}}
\begin{table}[t]
  \caption{Preparation steps covered by \sysname.}
  \label{tab:feature_coverage}
  
  \centering
  \footnotesize
  \setlength{\tabcolsep}{2pt}
  \renewcommand{\arraystretch}{1.06}
  \begin{tabular*}{\columnwidth}{@{\extracolsep{\fill}} l c l c @{}}
    \toprule
    \textbf{Type} & \textbf{Covered} & \textbf{Type} & \textbf{Covered} \\
    \midrule
    Join \& union                & \cmark & Derive column                   & \cmark \\
    Filter rows                  & \cmark & Fill empty cells                & \cmark \\
    Sort                         & \cmark & Edit \& replace cell data       & \cmark \\
    Deduplicate data             & \cmark & Normalize strings               & \cmark \\
    Pivot \& unpivot             & \cmark & Detect \& change encoding       & \xmark \\
    Split column                 & \cmark & Change column data type         & \cmark \\
    Merge columns                & \cmark & Assign semantic data type       & \xmark \\
    Delete empty \& invalid rows & \cmark & Discover \& merge external data & \xmark \\
    Delete column                & \cmark & Generate primary key            & \cmark \\
    Rename column                & \cmark & \multicolumn{2}{c}{} \\
    \bottomrule
  \end{tabular*}
  
\end{table}
}

\begin{figure*}[t]
  \centering
  \includegraphics[width=0.95\textwidth]{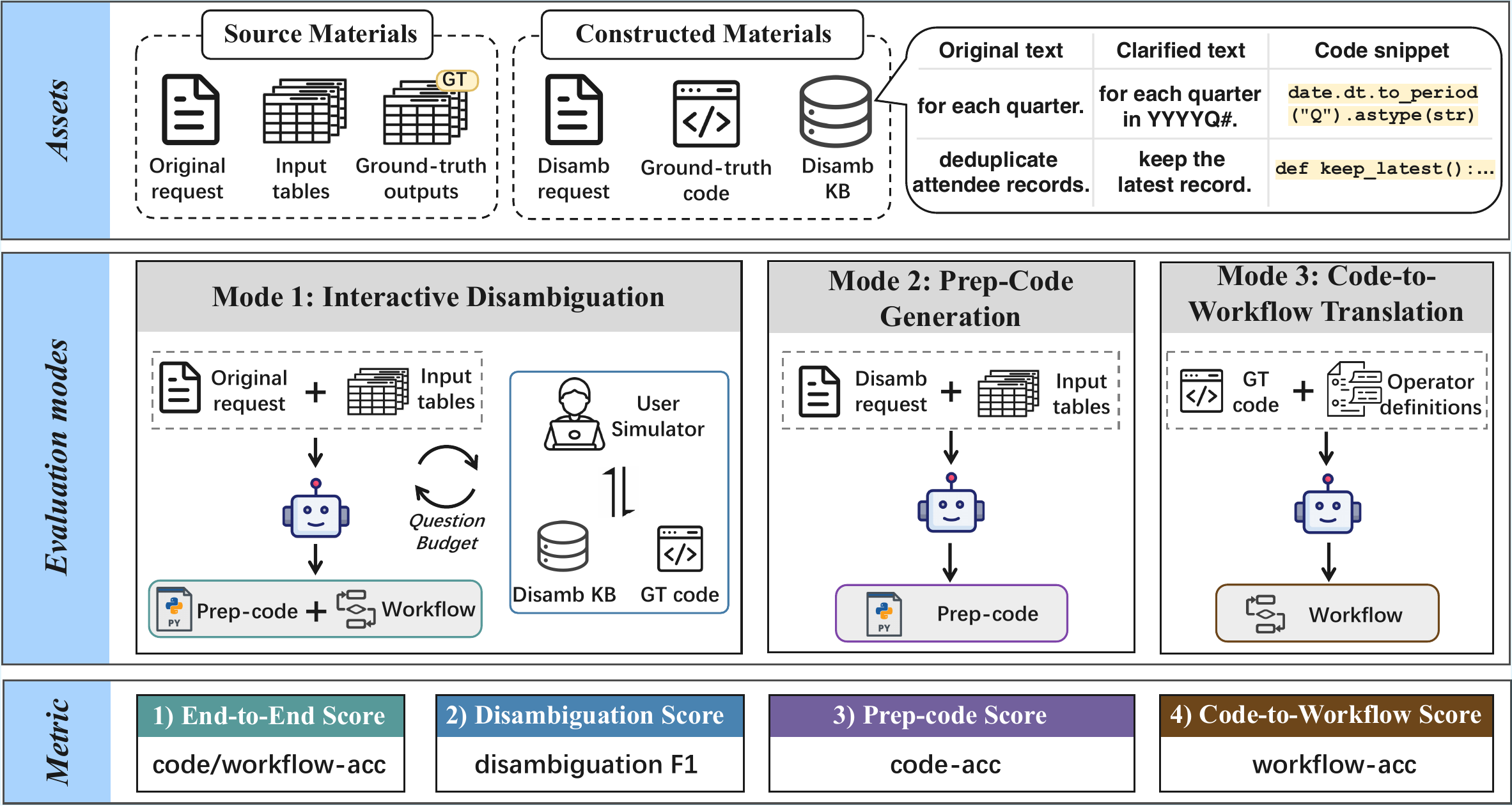}
  \caption{Evaluation pipeline in \sysname. All modes share the same input tables, reference outputs, and comparator; varying the input assets isolates different capabilities.}
  \Description{Pipeline diagram showing assets, three evaluation modes, and score components.}
  \label{fig:eval_pipeline}
\end{figure*}

\section{Capability Evaluation and Analysis}
\label{sec:capability}

Figure~\ref{fig:eval_pipeline} provides an overview of \sysname. 
In terms of benchmark assets, \sysname consists of source materials and constructed materials. 
Source materials are extracted from Preppin\textquotesingle{} Data, including \textit{original request}, \textit{input tables}, and \textit{ground-truth outputs}. 
We construct four new materials (detailed in Section~\ref{sec:benchmark:construction}), including \gtcode, \dreq, \dkb and \gtflow.
In terms of evaluation modes, \sysname consists of three modes.  
We explain their details in the rest of this section.

\subsection{Interactive Disambiguation}
\label{sec:interactive_disambiguation}

User requests in NL-driven data preparation are often ambiguous, resulting in multiple valid interpretations.
This section presents how \sysname evaluates the \emph{interactive-disambiguation} capability of an agent system. 

\noindent\textbf{Evaluation Mode.}
As shown in Mode~1 of \figref{fig:eval_pipeline}, \sysname provides $\oreq$, $\intab$, and an LLM-based \emph{user simulator}.
An agent system may inspect $\intab$ and interact with the user simulator to clarify ambiguities in the $\oreq$, subject to a question budget.
After the interaction, the agent generates prep-code and then translates it into a workflow, following the procedures as described in \secref{sec:prep_code_generation} and \secref{sec:code_to_gui_translation}.

\noindent\underline{User simulator.}
When an agent identifies ambiguities in an \oreq, it ideally relies on a human user to answer clarification questions.
To automate this process, recent work has adopted LLM-based user simulators that provide natural-language feedback~\cite{wang2023mint}.
Such a simulator takes a clarification question posed by the agent as input and generates a response intended to mimic a human user.
However, unconstrained simulators may leak information from the reference solution or deviate from the task objectives~\cite{huo2025bird}.
To address this issue, we define a response policy as follows.

A question is considered \emph{invalid} if it cannot be mapped to a specific ambiguity entry, such as requesting the reference output or the full solution.
For invalid questions, the user simulator returns a rejection response with an explanation.
The agent is informed of the invalid-question definition before evaluation.
For a valid question, if it matches an ambiguity entry in \dkb, the simulator returns the corresponding resolution in natural language.
Otherwise, the simulator provides a response based on \gtcode, staying strictly within the scope of the question.
For example, it rejects invalid questions such as ``Can you provide the prepared table ?'' (Figure~\ref{fig:intro}(c)). 
For valid questions, it either returns a natural-language resolution referencing a \dkb item (e.g., ``deduplicate: keep the latest record'', Figure~\ref{fig:eval_pipeline}) or provides scoped information from \gtcode (e.g., ``quarter uses \texttt{YYYYQ\#} like \texttt{2024Q1}'').


\noindent\underline{Evaluation Metrics}.
While Mode~1 (\figref{fig:eval_pipeline}) is designed for evaluating disambiguation quality, it also plays an important role in measuring end-to-end performance for NL-driven data preparation.
We report end-to-end performance using \emph{code-acc} and \emph{workflow-acc}, and evaluate disambiguation quality using \emph{disambiguation F1}.

\sloppy 

We compute \emph{code-acc} and \emph{workflow-acc} by executing the generated prep-code or workflow on $\intab$ to obtain an output table $\hat{T}_i$, and checking whether it matches $\gtout$ $T_i$ with a \emph{comparator}. 
The \emph{comparator} aligns rows in $\hat{T}_i$ and $T_i$ by a primary key, which we label for each task using a script and manually verify. When no single column uniquely identifies rows, we use a composite key.
After alignment, the \emph{comparator} compares the remaining columns with light normalization on date formats, numeric precision, and whitespace.
The \emph{comparator} is invariant to row order unless explicitly required by the request. For tasks with multiple output tables, all tables need to match.
Formally, $\textsc{Acc}=\frac{1}{N}\sum_{i=1}^{N}\mathbf{1}[\textsc{Compare}(\hat{T}_i,T_i)]$.

\fussy

We use \emph{disambiguation F1} to measure disambiguation quality.
Let $A$ be the set of \dkb entries for a task.
Each valid question can match at most one entry in $A$.
Let $M \subseteq A$ be the subset of entries that are matched at least once, with each entry counted only once.
Let $k$ be the total number of questions.
We define $\textsc{Precision}=|M|/k$ and $\textsc{Recall}=|M|/|A|$, and report their harmonic mean as \emph{disambiguation F1}.
Redundant or off-target questions increase $k$ without enlarging $M$, thereby lowering precision.

\noindent\underline{Capability Requirements}.
To operate in Mode~1, an agent system should support asking clarification questions before code generation.
At each round, the agent receives $\oreq$, $\intab$, the clarification history, and the remaining question budget.
It then chooses one of two actions: asking a clarification question to the simulator, or stopping the interaction and proceeding to code generation.
We set the budget to $\lceil 2.5\cdot |A| \rceil$ by default.
Since the agent does not know $|A|$, the budget should exceed $|A|$ to leave room for redundant or off-target questions. Yet it should not be too large, or it would let agents succeed by asking questions indiscriminately. The $2.5\times$ multiplier balances the two.
The interaction ends when the agent determines that disambiguation is complete, or when the question budget is exhausted.

\begin{figure}[t]
  \centering
  \includegraphics[width=0.95\columnwidth]{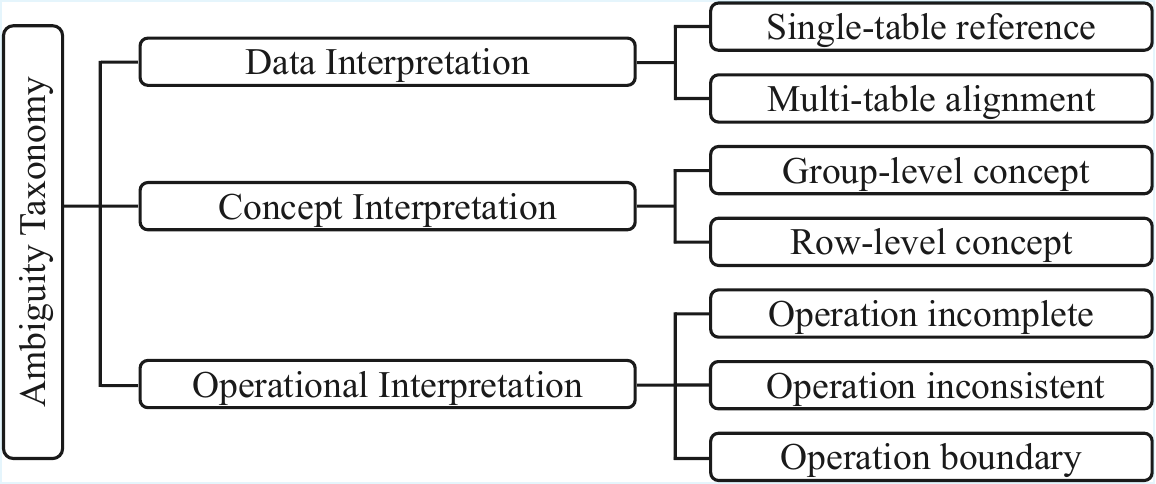}
  \caption{Ambiguity taxonomy in \sysname.}
  \Description{Sunburst-style taxonomy grouping data-preparation ambiguities into data interpretation, concept interpretation, and operational interpretation.}
  \label{fig:amb_tax}
\end{figure}

\noindent\textbf{Taxonomy and Statistical Analysis.} To better understand ambiguities in data preparation tasks, we present a taxonomy of common ambiguity types in user requests.
We observe that a request may be ambiguous for three main reasons:
(i) \emph{Data interpretation}: the request does not clearly specify which part of the data an operation should apply to;
(ii) \emph{Concept interpretation}: the request refers to concepts that are not well defined;
(iii) \emph{Operational interpretation}: the request does not clearly specify how to handle edge cases for certain operations.
\figref{fig:amb_tax} illustrates this taxonomy.

\noindent\underline{(i) Data Interpretation}.  
We further identify two common subtypes under this category.
\emph{Single-table reference} occurs when an entity mentioned in the request cannot be uniquely linked to a specific column or value within a table (e.g., ``sales'' could refer to both \texttt{sales\_\allowbreak amount} and \texttt{sales\_\allowbreak count}).
\emph{Multi-table alignment} arises when a request involves multiple tables but does not specify which fields should be used to align them.

\noindent\underline{(ii) Concept Interpretation}.  We observe two common cases.
\emph{Group-level concept} refers to quantities defined over a group of rows, but the aggregation logic is unclear.
For example, ``average price'' may refer to a simple average or a weighted average, and ``monthly revenue'' may be computed based on different date fields.
\emph{Row-level concept} refers to conditions or labels applied to individual rows, but their definitions are underspecified.
For example, terms such as ``recent records,'' ``high-value customers,'' or ``valid entries'' lack clear thresholds or criteria.

\noindent\underline{(iii) Operational Interpretation}.  
Consider a shipping-fee rule: fragile items cost \$5, and orders of \$100+ cost \$3.  
This example illustrates three common types of operational ambiguity.
\emph{Operation incomplete} occurs when an order matches none of the rules.
For example, a non-fragile order under \$100 is not covered by either rule, and the request does not specify what fee to apply.
\emph{Operation inconsistent} occurs when an order matches multiple rules.
For example, a fragile order over \$100 satisfies both rules, but it is unclear whether the fee should be \$5, \$3, or the sum of both fees.
\emph{Operation boundary} occurs when threshold conditions are unclear.
For example, it may be ambiguous whether an order of exactly \$100 qualifies as ``\$100+.''

\begin{figure}[!t]
  \centering
  \includegraphics[width=0.95\columnwidth]{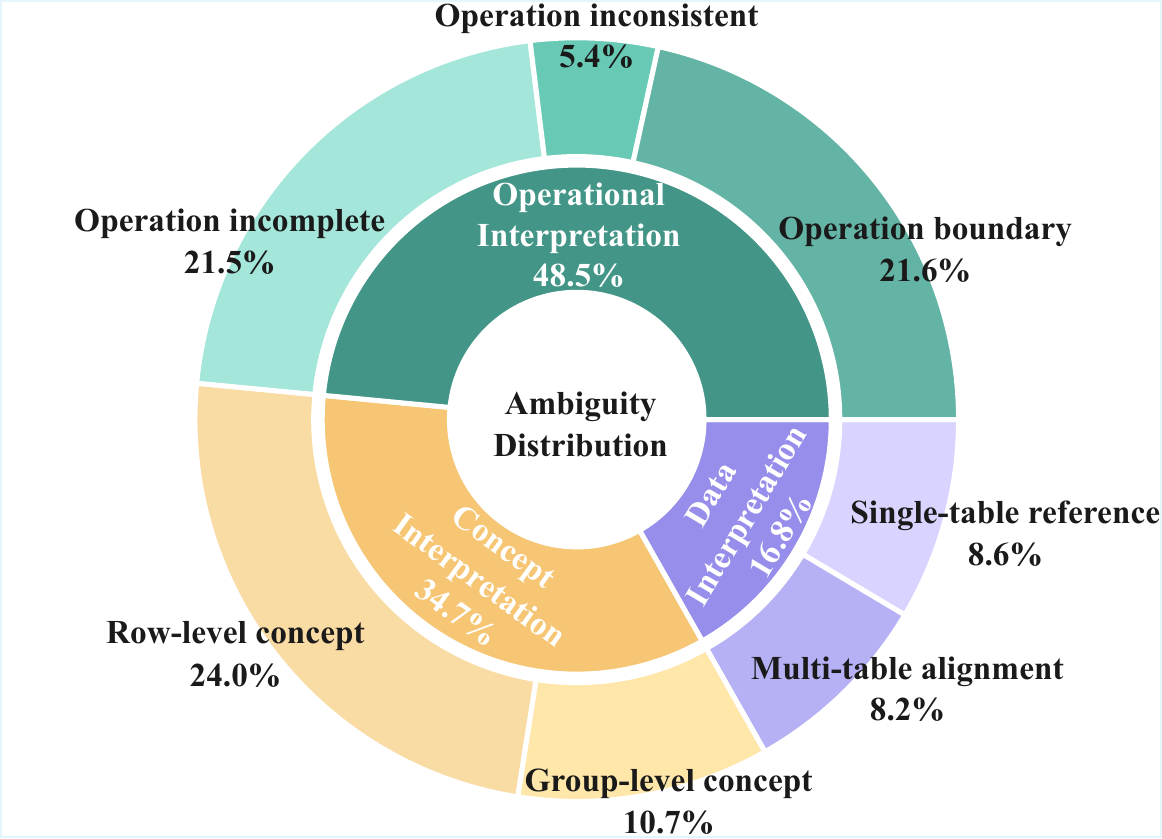}
  \caption{Distribution of ambiguities in \sysname.}
  \Description{Donut chart showing the percentage distribution of ambiguity categories and subtypes in the benchmark.}
  \label{fig:amb_distribution}
\end{figure}

\noindent\underline{Statistics}.  
\figref{fig:amb_distribution} reports the distribution of ambiguity types in \sysname.
\emph{Operational interpretation} is the most common, accounting for 48.5\% of all ambiguities.
This suggests that users often leave edge-case handling unspecified.
In contrast, \emph{data interpretation} accounts for only 16.8\% of ambiguities, but mistakes in this category often have a broad impact.
For example, an incorrect column reference or join key can affect all subsequent steps.
These results indicate that resolving data references early is important to avoid downstream errors, and that systems should actively encourage users to clarify edge-case handling.

\subsection{Prep-Code Generation}  
\label{sec:prep_code_generation}  
After disambiguation clarifies the intended operations, the system generates executable \emph{prep-code} for the input tables. However, disambiguation alone is not sufficient.
In practice, real-world tables often contain irregular or unexpected values.
These irregularities may not appear in small table samples or may be unknown to the user at the time of specification.  
As a result, they are not mentioned in the request, but can still lead to incorrect results or execution failures if not properly handled.

\noindent\textbf{Evaluation Mode.}
\label{sec:prep_code_generation:eval}
Mode~2 (\figref{fig:eval_pipeline}) shows how to evaluate prep-code generation in isolation, without requiring disambiguation.  
\sysname provides \dreq (instead of \oreq) and \intab, and the correctness is measured by \emph{code-acc}.

\noindent\underline{Capability Requirements}.
Mode~2 evaluates whether an agent can generate correct prep-code under practical table constraints. That is,
table samples may miss irregular values in full tables, while  full tables may exceed LLM context limits or make it harder for the LLM to focus on relevant details.
Moreover, different serialization formats can lead to varying model behavior, and no single format is consistently best~\cite{sui2024table}.

To succeed under Mode~2, an agent needs to reason about properties of the full tables without directly putting them into a prompt.
We therefore require the agent to support data profiling. One possible profiling procedure is as follows.
Before generating prep-code, the agent uses \dreq and a table sample to generate profiling code that scans the full tables.
If the initial profiling result is insufficient, the agent may revise and rerun the profiling once.
The profiling results are summarized and provided as additional input for prep-code generation.

\noindent\textbf{Taxonomy and Statistical Analysis.}
\label{sec:prep_code_generation:dataset}
To better understand the challenges of Mode~2, we analyze the data irregularity patterns within \sysname. Through a manual inspection of $\intab$ alongside $\gtcode$, we identified 112 tasks (constituting 36.6\% of \sysname) that exhibit irregularities in the raw inputs.
A task may involve multiple patterns.  
\tabref{tab:irregularity-taxonomy} presents a taxonomy of patterns, categorized into \emph{structural issues} and \emph{cell-level variants}.

Structural issues affect table schema integrity. (i) \emph{Header irregularities} refer to problematic header definitions (e.g., missing/duplicate column names). (ii) \emph{Row irregularities} denote extraneous rows within the table body, such as notes or interleaved headers. These two issues are less common but can break downstream processing if not detected early.

Cell-level variants are more common. 
(i) \emph{Format variants} refer to values of the same type expressed in different formats, such as dates written as \texttt{2020-01-02} and \texttt{09/04/2021} in the same column.  
(ii) \emph{Spelling variants} refer to the same entity appearing under different spellings, such as \texttt{Jagadish} and \texttt{JAGADISH}.  
(iii) \emph{Missing variants} refer to inconsistent null-value representation, such as empty strings, \texttt{N/A}, or \texttt{null}.  
These issues are easy to overlook in a small sample, which can lead to wrong grouping, filtering, and aggregation results.

\begin{table}[t]
\centering
\small
\setlength{\tabcolsep}{2pt}
\caption{Data irregularity patterns in \sysname.}
\label{tab:irregularity-taxonomy}

\begin{tabularx}{\linewidth}{@{}lXr@{}}
\toprule
\textbf{Pattern} & \textbf{Definition} & \textbf{Percentage} \\
\midrule
Header Irregularities & Problematic header definitions & 7.1\% \\
\addlinespace[0.4em]
Row Irregularities & The table contains invalid rows & 15.2\% \\
\addlinespace[0.4em]
Format Variants & Diverse formats for the same data type & 34.0\% \\
\addlinespace[0.4em]
Spelling Variants & Diverse entity spellings & 20.5\% \\
\addlinespace[0.4em]
Missing Variants & Inconsistent null-value representation & 33.0\% \\
\bottomrule
\end{tabularx}
\end{table}
    
\subsection{Code-to-Workflow Translation}
\label{sec:code_to_gui_translation}

Executable code can be difficult for non-technical users to inspect or modify.  
We therefore study \emph{code-to-workflow translation}, which converts prep-code into an equivalent operator workflow.
The workflow is a JSON-serialized DAG, where each node corresponds to a single preparation step with its parameters, making the logic easier to understand, check, and refine.

\noindent\textbf{Design Rationale.} 
Visual workflows are widely used to help end users inspect and edit data preparation logic. Commercial systems such as Tableau Prep~\cite{tableau_prep_about} and SAS Data Preparation~\cite{sas_data_prep} are built around this paradigm, and prior studies on interactive data transformation provide further support~\cite{kandel2011wrangler,santos2025interactive}.  
In \sysname, workflows are designed to support both execution and interpretation.
They can be executed on the input tables to produce the output, while exposing intermediate tables at operator nodes to help users localize transformation errors.

\noindent\textbf{Evaluation Mode.}
Mode~3 (\figref{fig:eval_pipeline}) shows how to evaluate code-to-workflow translation in isolation.
\sysname provides \gtcode and a fixed set of workflow operator definitions.
For each task, all agents are given the same \gtcode.
Each agent is required to output an executable workflow that preserves the semantics of \gtcode.

\noindent\underline{Evaluation Metrics}.  
Mode~3 (\figref{fig:eval_pipeline}) evaluates code-to-workflow translation in isolation using \emph{workflow-acc}.  
Each translated workflow is executed on $\intab$ and classified into one of three outcomes:  
(i)~\emph{Fail}, if the workflow violates format or operator specifications and cannot be executed;  
(ii)~\emph{Wrong}, if it executes but produces an output different from $\gtout$; or  
(iii)~\emph{Correct}, if it executes and produces an output that matches $\gtout$.  
\emph{workflow-acc} is the fraction of workflows classified as \emph{Correct}.

\noindent\underline{Capability Requirements}.
Mode~3 evaluates two capabilities.  
The agent is expected to follow the workflow operator definitions, including their semantics and parameter constraints.
It should also apply these operators to translate \gtcode into an operator DAG that preserves the semantics of the original code.

\begin{table}[t]
\centering
\small
\setlength{\tabcolsep}{4pt}
\renewcommand{\arraystretch}{1.1}
\caption{Operators and their frequency in \sysname.}
\label{tab:operator_vocab}
\begin{tabularx}{\columnwidth}{@{}l >{\raggedright\arraybackslash}X r@{}}
\toprule
\textbf{Operator} & \textbf{Description} & \textbf{Frequency} \\
\midrule
\textsc{Input}     & Load table from file                                          & 100\% \\
\textsc{Join}      & Join tables on key columns                                     & 70.3\% \\
\textsc{Union}     & Append tables by column names                                  & 35.0\% \\
\textsc{Project}   & Select, rename, cast, or derive columns                         & 100\% \\
\textsc{Filter}    & Select rows by predicate                                       & 64.4\% \\
\textsc{Aggregate} & Group by keys and compute aggregates                           & 62.8\% \\
\textsc{Sort}      & Order rows by keys                                             & 69.3\% \\
\textsc{Pivot}     & Reshape between wide and long layouts                          & 25.8\% \\
\textsc{Dedup}     & Select one row per key                                         & 24.8\% \\
\textsc{Output}    & Write table to file                                            & 100\% \\
\textsc{Script}    & Inline code fallback                                           & 14.4\% \\
\bottomrule
\end{tabularx}
\end{table}

\noindent\textbf{Operators and Statistical Analysis.}
To better understand the challenges of code-to-workflow translation, we define a workflow representation and an execution engine, and analyze the frequency of operator usage across tasks in \sysname.

\noindent\underline{Operator Design}.
Our operator set is inspired by Tableau Prep and covers common steps in visual data preparation.  
Each node in the workflow represents a single atomic transformation with its parameters.  
When a GUI step involves multiple actions, we decompose it into separate nodes.  
For logic that is hard to express with operators (e.g., complex string parsing or custom date handling), we provide a \textsc{Script} node as a fallback, with code limited to 1{,}500 characters.

\noindent\underline{Execution Engine}.  
The execution engine first checks whether the workflow has a valid structure and conforms to the operator specifications.
It then executes the workflow in topological order, creating intermediate tables as needed and discarding them once they are no longer used.  
The engine also provides optional tracing and error messages to support debugging.

\noindent\underline{Operator Overview.}
We define a total of 11 workflow operators in \sysname.
\tabref{tab:operator_vocab} lists these operators and their descriptions. 
It also reports the frequency of each operator across all tasks to show operator usage in the benchmark. 
From the table, we observe that \textsc{Pivot} and \textsc{Dedup} each appears in about 25\% of the tasks, highlighting the significance of reshaping and record-level resolution in data preparation workflows.
The \textsc{Script} operator appears in 14.4\% of the tasks, and is used when the workflow includes logic that is hard to express with standard operators, such as complex parsing or custom date handling. 
In comparison, \textsc{Pivot}, \textsc{Dedup}, and \textsc{Script} are less common in Text-to-SQL QA benchmarks, highlighting the distinct characteristics and additional challenges of \sysname.


\noindent\textbf{User-Centric Verification Study.}
Workflow-acc evaluates whether a translated workflow is correct, but it does not measure whether users can inspect the transformation logic. We conducted a verification study with 16 participants who had experience with data preparation. The study used six tasks drawn from \sysname cases, including four incorrect and two correct candidate implementations. None of the participants was involved in constructing \sysname.

Each task included a transformation goal, an input preview, and one candidate implementation, shown either as code or as a workflow. The workflow presented each step as an operator card with its purpose and parameters. 
In a counterbalanced design, each participant inspected three code tasks and three workflow tasks, and each task appeared in both formats across participants. Participants judged correctness, reported confidence, and answered a final preference question.

\tabref{tab:user_study} reports the results. Workflow yielded higher judgment accuracy than code, increasing from 72.9\% to 85.4\%. Participants also reported higher inspection confidence with workflow, increasing from 3.6 to 4.1 on a 1--5 scale.  
In the final preference question, 9 of 16 participants preferred workflow for inspection, 4 preferred code, and 3 reported no clear preference.
These results provide preliminary user-centric evidence that workflow representations help users inspect data-preparation logic by exposing operator boundaries and parameters.

\begin{table}[!t]
\centering
\begingroup

\small
\setlength{\tabcolsep}{5pt}
\renewcommand{\arraystretch}{1.1}

\caption{User-centric verification study.}
\label{tab:user_study}

\begin{tabular*}{\linewidth}{@{\extracolsep{\fill}}lcc@{}}
\toprule
\textbf{Metric} & \textbf{Code} & \textbf{Workflow} \\
\midrule
Judgment accuracy & 72.9\% & 85.4\% \\
Inspection confidence (1--5) & 3.6 & 4.1 \\
Preferred for inspection & 4/16 & 9/16 \\
\bottomrule
\end{tabular*}

\arrayrulecolor{black}
\endgroup
\end{table}






\section{Experiments}
\label{sec:experiments}

In this section, we evaluate LLM-based agents on \sysname and
analyze the results, aiming to answer the following questions:

{\small
\begin{itemize}[leftmargin=*,nosep]
  \item \textbf{RQ\,1}: Are agents ready for end-to-end NL-driven data preparation?
  \item \textbf{RQ\,2}: Does NL ambiguity significantly hurt performance?
  \item \textbf{RQ\,3}: Can agents effectively resolve ambiguity through interaction?
  \item \textbf{RQ\,4}: Can agents accurately generate prep-code from unambiguous NL?
  \item \textbf{RQ\,5}: Can agents reliably translate prep-code into GUI workflows?
\end{itemize}
}

\begin{table*}[t]
  \centering\small
  \renewcommand{\arraystretch}{1.25}
  \newcolumntype{C}[1]{>{\centering\arraybackslash}p{#1}}
  \newcolumntype{M}[1]{>{\centering\arraybackslash}m{#1}}
  \caption{RQ1 results across five evaluation settings. Best and second-best in each column are in bold and underlined, respectively.}
  \label{tab:rq1-main}
  \resizebox{0.95\textwidth}{!}{%
  \begin{tabular}{l !{\vrule} C{1.1cm}C{1.1cm} !{\vrule} C{1.1cm}C{1.1cm} !{\vrule} C{1.1cm}C{1.1cm} !{\vrule} C{1.1cm}C{1.1cm} !{\vrule} C{1.1cm}C{1.1cm}}
    \noalign{\hrule height 1.2pt}
    \multirow{3}{*}[-1.8ex]{\textbf{Model}}
    & \multicolumn{4}{c!{\vrule}}{\textbf{End-to-End}}
    & \multicolumn{6}{c}{\textbf{Isolated Capabilities}} \\
    \cline{2-5}\cline{6-11}
    & \multicolumn{2}{c!{\vrule}}{\makecell{\textbf{Prep-Code}}}
    & \multicolumn{2}{c!{\vrule}}{\makecell{\textbf{GUI Workflow}}}
    & \multicolumn{2}{c!{\vrule}}{\makecell{\textbf{Interactive}\\\textbf{Disambiguation}}}
    & \multicolumn{2}{c!{\vrule}}{\makecell{\textbf{Prep-Code}\\\textbf{Generation}}}
    & \multicolumn{2}{c}{\makecell{\textbf{Code-to-Workflow}\\\textbf{Translation}}} \\
    & Acc. & Cost & Acc. & Cost & F1 Score & Cost & Acc. & Cost & Acc. & Cost \\
    \hline
    \multicolumn{11}{l}{\textit{Proprietary Models}} \\
    \hline
    GPT-5.1-Codex     & \textbf{54.9} & 115.40 & \textbf{34.6} & 264.10 & \textbf{51.4} & 74.82 & \textbf{85.3} & 82.74 & \textbf{67.7} & 159.10 \\
    Claude-Sonnet-4.5 & 52.0          & 114.00 & 24.5          & 223.20 & \underline{51.3} & 66.87 & 70.3 & 70.12 & 44.8 & 102.30 \\
    Gemini 3 Flash    & \underline{53.3} & 21.40 & 22.2          & 41.66  & 51.0 & 10.67 & \underline{74.8} & 11.27 & \underline{52.3} & 16.48 \\
    Grok Code Fast 1  & 30.1          & 18.61  & 13.1          & 35.56  & 34.5 & 13.25 & 57.5 & 10.66 & 45.1 & 20.46 \\
    GPT-4o            & 16.7          & 105.40 & 5.2           & 141.70 & 44.6 & 42.17 & 39.5 & 43.28 & 19.9 & 61.89 \\
    \hline

    \multicolumn{11}{l}{\textit{Open-Weight Models}} \\
    \hline
    Kimi K2 Thinking  & 49.7 & 36.04 & \underline{30.1} & 75.04 & 45.9 & 18.29 & 69.3 & 23.68 & 46.1 & 18.33 \\
    GLM-4.7           & 41.5 & 24.74 & 19.9 & 53.88 & 48.5 & 14.74 & 65.7 & 16.70 & 34.6 & 19.74 \\
    Qwen3-235B-A22B   & 38.2 & 36.99 & 19.3 & 47.51 & 41.3 & 26.68 & 67.7 & 16.76 & 35.0 & 22.29 \\
    DeepSeek-V3.2     & 44.8 & \underline{6.62} & 15.7 & \underline{11.59} & 45.7 & \underline{4.48} & 65.4 & \underline{3.40} & 27.8 & \underline{4.85} \\
    DevStral 2        & 33.3 & \textbf{1.89} & 8.5  & \textbf{3.16} & 42.3 & \textbf{1.24} & 44.1 & \textbf{1.06} & 10.5 & \textbf{1.82} \\
    \noalign{\hrule height 1.2pt}
  \end{tabular}%
  }
\end{table*}

\subsection{Experimental Settings}
\label{sec:experiments:setup}

We systematically evaluated agents for NL-driven data preparation, covering end-to-end performance as well as three core capabilities: interactive disambiguation, prep-code generation, and code-to-workflow translation.

\noindent\textbf{Agent Actions.}
In \sysname, an LLM-based agent can take four actions:
(1)~\emph{Clarify}: ask clarification questions, subject to a budget $K$;
(2) \emph{Profile}: inspect input tables by generating profiling code;
(3) \emph{Code}: generate data preparation code;
(4) \emph{Translate}: convert code into an executable GUI workflow.
For \emph{Profile}, \emph{Code}, and \emph{Translate}, the system executes the generated artifacts and returns execution results or an error.
\emph{Profile} allows up to 2 attempts, while \emph{Code} and \emph{Translate} allow up to 3. \emph{Profile} does not enter evaluation directly. The agent can still generate code from the sample if profiling fails. \emph{Code} and \emph{Translate} produce the evaluated artifacts, so they need more retries to recover from execution errors.

\noindent\textbf{Models.}
We evaluated ten models: five proprietary (GPT-5.1-Codex, Claude-Sonnet-4.5, Gemini 3 Flash, Grok Code Fast 1, GPT-4o) and five open-weight (Kimi K2 Thinking, GLM-4.7, Qwen3-235B-A22B, DeepSeek-V3.2, DevStral 2).
All models were run with temperature $\tau=0.7$.
For interactive evaluation, we used DeepSeek-V3.2 ($\tau=0$) as the user simulator to answer clarification questions.

\noindent\textbf{Metrics.}
We report accuracy via \emph{code-acc} and \emph{workflow-acc}, and disambiguation quality via \emph{disambiguation F1} (\secref{sec:interactive_disambiguation}).
To assess how well agents resolved different ambiguity types, we report \emph{type-level recall} (i.e., entry coverage) for each ambiguity category.
For a type $t$, let $A_t$ be the set of \dkb entries labeled as $t$, and let $M_t \subseteq A_t$ be the entries matched by at least one valid question, with each entry counted at most once.
We focused on recall, as unmatched questions cannot be assigned to a specific ambiguity type and thus do not support per-type precision.
We define $\textsc{Recall}(t)=|M_t|/|A_t|$.
We also reported the average cost per task in USD ($\times 10^{-3}$). 
Cost is computed over the execution path used by each evaluation setting, including all LLM calls and retries. For example, Interactive Disambiguation includes only \emph{Clarify}, while End-to-End Prep-Code additionally includes \emph{Profile} and \emph{Code}.

\begin{figure}[t]%
  \centering
  \makebox[\columnwidth][c]{\includegraphics[width=0.9\columnwidth]{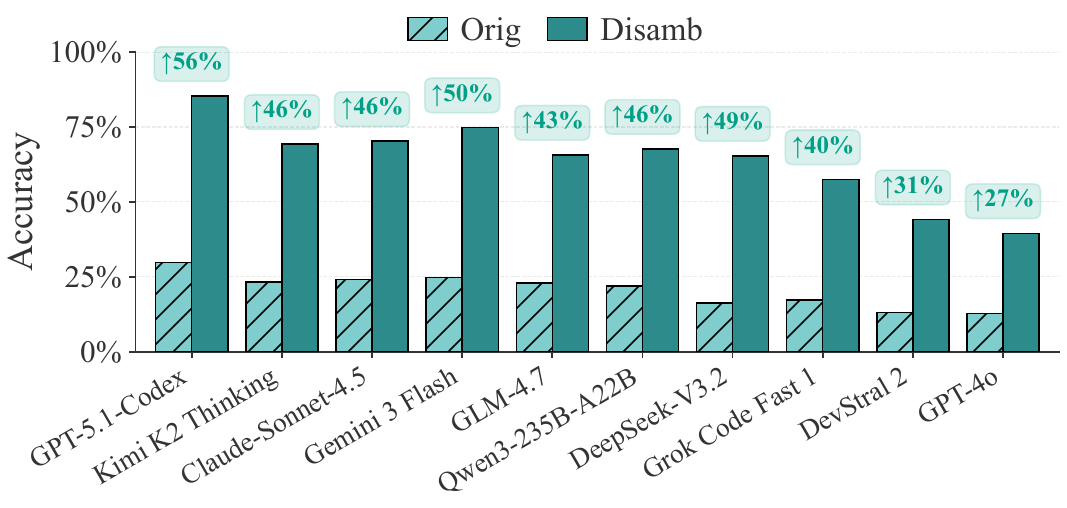}}
  \caption{Ambiguity gap across agents.}
  \Description{Bar chart comparing agent performance with original requests and disambiguated requests.}
  \label{fig:rq2_overall}
\end{figure}

\subsection{RQ1: End-to-End Performance}
\label{sec:experiments:rq1}
RQ1 asks whether current agents are ready for end-to-end NL-driven data preparation. 
\tabref{tab:rq1-main} reports performance across five evaluation settings.  
The first two followed the end-to-end protocol of Mode~1 (\secref{sec:interactive_disambiguation}):  
the agent received $\oreq$ and could perform \emph{Clarify}, \emph{Profile}, \emph{Code}, and \emph{Translate} actions.
We reported task accuracy for both prep-code and GUI workflows.
The remaining three capabilities were evaluated in isolation. 
Interactive Disambiguation was evaluated in Mode~1 using disambiguation F1. 
Prep-Code Generation was evaluated in Mode~2 (\secref{sec:prep_code_generation}), where the agent received $\dreq$ and only performed \emph{Profile} and \emph{Code};
Code-to-workflow Translation was evaluated in Mode~3 (\secref{sec:code_to_gui_translation}), where the agent received $\gtcode$ and only performed \emph{Translate}.
For brevity, we refer to these five settings as E2E-Workflow, E2E-Code, Disambig, CodeGen, and Translation.

\begin{figure}[t]%
  \centering  \includegraphics[width=0.9\columnwidth]{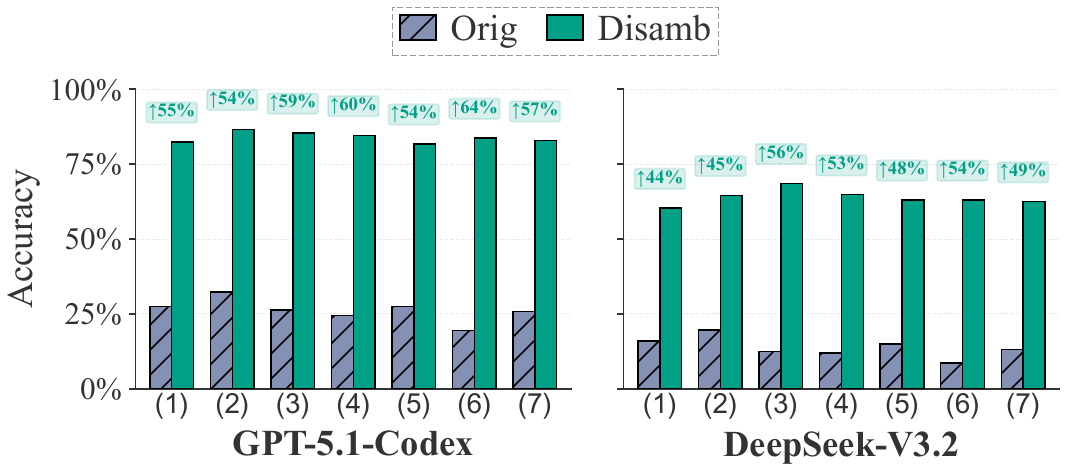}
  \caption{Ambiguity gap by type: (1) single-table reference, (2) multi-table alignment, (3) group-level concept, (4) row-level concept, (5) operation incomplete, (6) operation inconsistent, and (7) operation boundary.}
  \Description{Grouped chart showing accuracy gains from disambiguation across seven ambiguity types.}

  \label{fig:rq2_bytype}
\end{figure}

\noindent\textbf{Current agents are not ready for end-to-end NL-driven data preparation.}
As shown in the E2E-Code and E2E-Workflow accuracy columns of \tabref{tab:rq1-main}, even the best-performing model (GPT-5.1-Codex) achieved only 54.9\% on E2E-Code. Performance dropped substantially in the E2E-Workflow setting, where no model exceeded 35\% accuracy.

\noindent\textbf{Disambiguation leaves substantial room for improvement.}
As shown in Disambig (\tabref{tab:rq1-main}), disambiguation F1 scores ranged from 34.5 to 51.4, indicating that current models struggled to identify ambiguities in user requests.
Moreover, interactive disambiguation did not fully recover the accuracy achievable with a disambiguated request:
comparing E2E-Code with CodeGen, GPT-5.1-Codex dropped from 85.3\% to 54.9\%, and Gemini~3~Flash from 74.8\% to 53.3\%.
We examined this gap further in RQ3.

\noindent\textbf{Code-to-workflow translation remains the bottleneck.}
Comparing E2E-Code with E2E-Workflow, GPT-5.1-Codex dropped from 54.9\% to 34.6\%, reflecting errors introduced during translation.
Even with correct prep-code as input (Translation), accuracy remained capped at 67.7\%, confirming that workflow generation is a distinct and error-prone task.
This gap is structural: prep-code is free-form, while workflows need to conform to a schema-constrained, operator-based representation that models have not seen during training.
We analyze translation failures in RQ5.

\noindent\textbf{Cost-effectiveness varies across models.}
Higher cost does not guarantee better performance: GPT-4o is expensive yet performs poorly on E2E-Workflow.
In contrast, Gemini~3~Flash nearly matched GPT-5.1-Codex on E2E-Code (53.3\% vs.\ 54.9\%) at under one-fifth the cost.
These results highlight the need to consider both accuracy and cost in deployment decisions.

\begin{takeawaybox}
\textbf{Takeaway I.} Current agents are not ready for end-to-end NL-driven data preparation.
Performance is constrained by two bottlenecks: insufficient disambiguation limits prep-code accuracy, and workflow translation further degrades output.
\end{takeawaybox}

\begin{figure}[t]
  \centering
  \makebox[\columnwidth][c]{%
    \includegraphics[width=0.9\columnwidth]{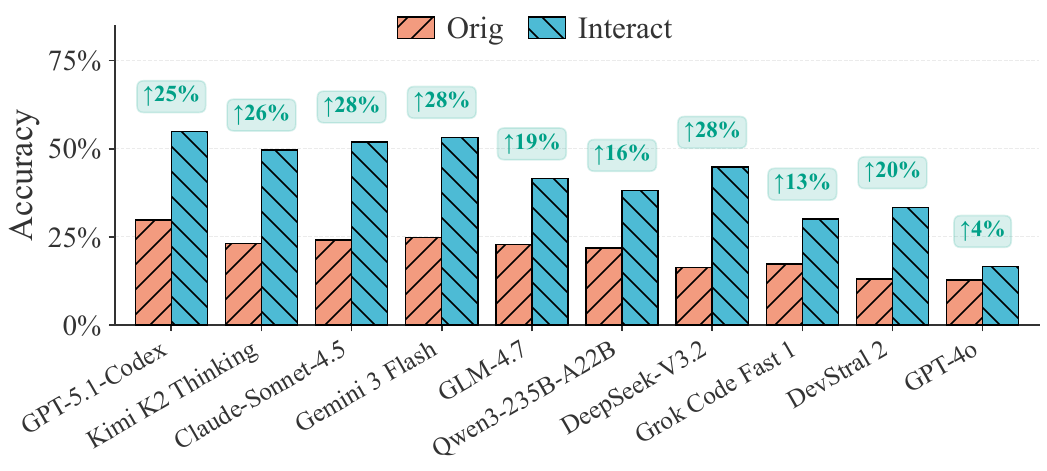}%
  }
  \caption{Interact gains across models.}
  \Description{Bar chart showing the gain from interactive clarification for each evaluated model.}
  \label{fig:rq3_interact}%
\end{figure}

\begin{figure}[t]%
  \centering
  \makebox[\columnwidth][c]{%
    \includegraphics[width=0.9\columnwidth]{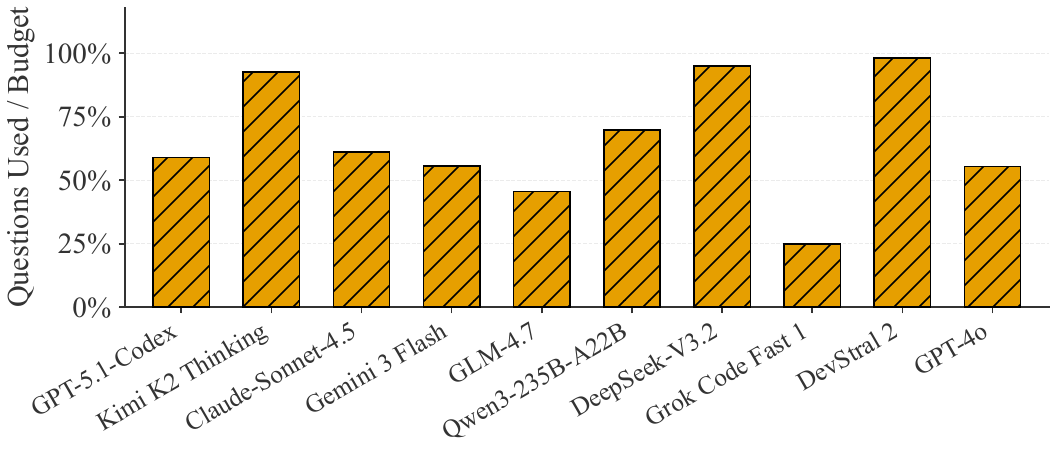}%
  }
  \caption{Question budget usage across models.}
  \Description{Bar chart comparing question-budget usage by model during interactive disambiguation.}
  \label{fig:rq3_budget}
\end{figure}

\subsection{RQ2: Impact of NL Ambiguity}
\label{sec:experiments:rq2}
RQ2 examines how NL ambiguity affects prep-code accuracy.
We compared two settings: \orig\ uses $\oreq$, while \disamb\ uses $\dreq$.
The agent received the request and could perform \emph{Profile} and \emph{Code} actions.
We define the \emph{ambiguity gap} as the accuracy increase from \orig\ to \disamb.

\noindent\textbf{NL ambiguity substantially degrades solution accuracy.}
Figure~\ref{fig:rq2_overall} shows a consistent ambiguity gap across all models, with improvements ranging from +27 to +56 points.
GPT-5.1-Codex achieved the highest \disamb\ accuracy (85\%) and the largest gap (+56\%), suggesting stronger models benefit more from disambiguation.
We examined whether interaction can close this gap in RQ3.

\noindent\textbf{Concept interpretation ambiguities show robust gains from disambiguation.}
Figure~\ref{fig:rq2_bytype} breaks down the ambiguity gap by type for GPT-5.1-Codex and DeepSeek-V3.2.
Both models show large and consistent gains on concept interpretation ambiguities, including group-level and row-level cases, where aggregation logic or thresholds need to be explicitly specified.
Operation inconsistent also improves substantially, as conflicting rules require explicit resolution.
Data interpretation ambiguities show smaller or more moderate gaps, as models can infer column mappings from schema.

\begin{takeawaybox}
\textbf{Takeaway II.} NL ambiguity substantially degrades accuracy, but the largest recoveries come from resolving ambiguities that determine transformation semantics rather than input interpretation alone.
\end{takeawaybox}



\begin{figure}[t]
  \centering
  \makebox[\columnwidth][c]{\includegraphics[width=0.9\columnwidth]{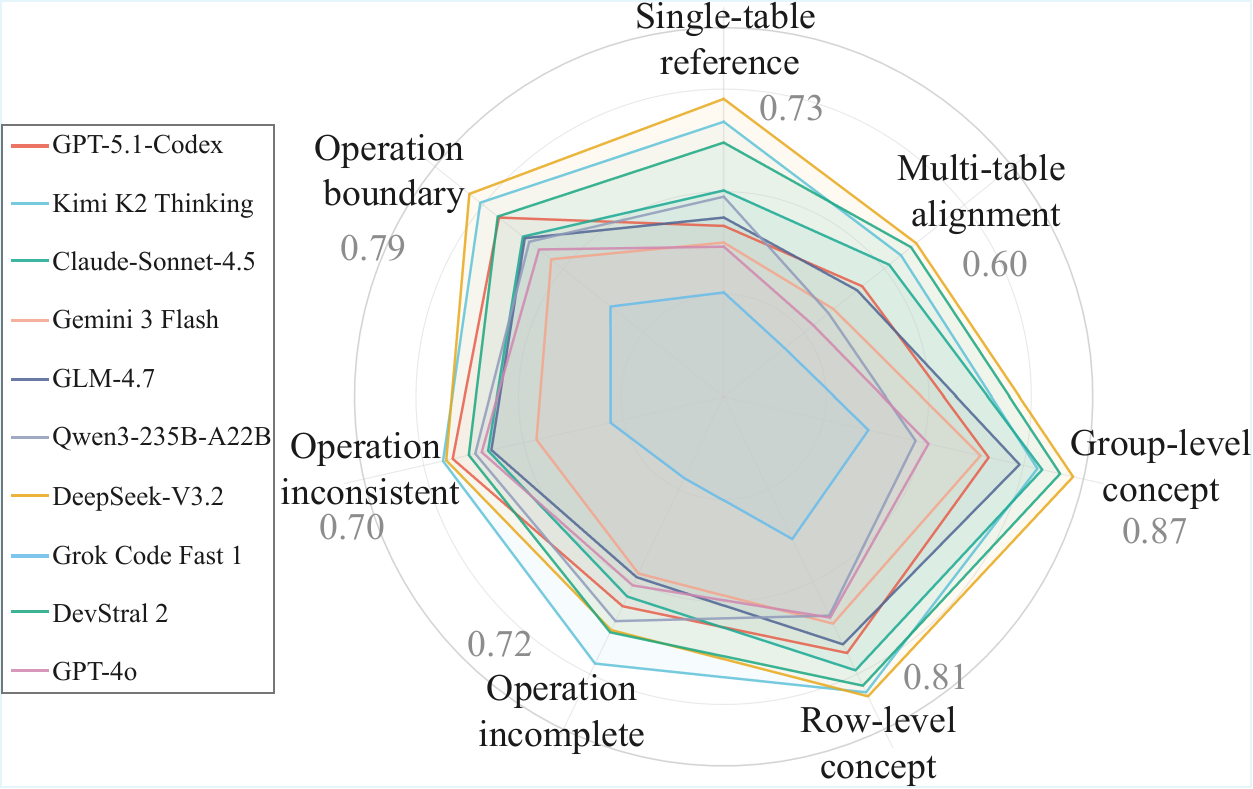}}
  \caption{Disambiguation recall by ambiguity type.}
  \Description{Chart reporting disambiguation recall for each ambiguity type.}
  \label{fig:rq3_bottleneck}
\end{figure}

\subsection{RQ3: Effectiveness of Interaction}
\label{sec:experiments:rq3}
The ambiguity gap in RQ2 raises a practical question: can agents resolve ambiguities through interaction?
We compared two settings: \disamb\ uses $\dreq$, while \interact\ uses $\oreq$.
In both settings, the agent could perform \emph{Profile} and \emph{Code} actions; \interact\ additionally permitted \emph{Clarify} actions to resolve ambiguities.
This setting simulated realistic usage, where the agent needed to identify ambiguities and resolve them through interaction.

\begin{figure}[t]
  \centering
  \makebox[\columnwidth][c]{\includegraphics[width=0.9\columnwidth]{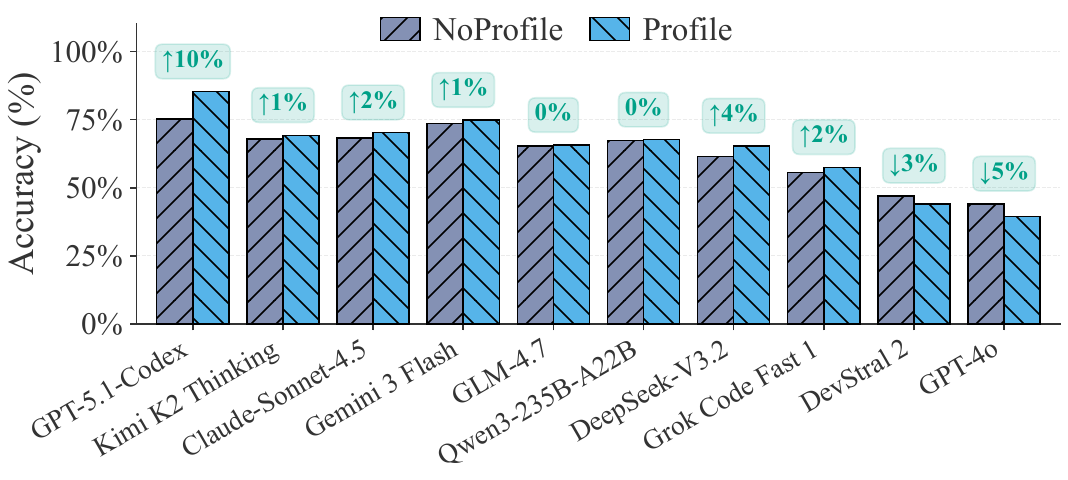}}
  \caption{Profiling gains across models.}
  \Description{Bar chart showing accuracy improvements from profiling across models.}
  \label{fig:rq4_overall}
\end{figure}

\begin{figure*}[t]
  \centering
  \includegraphics[width=0.95\textwidth]{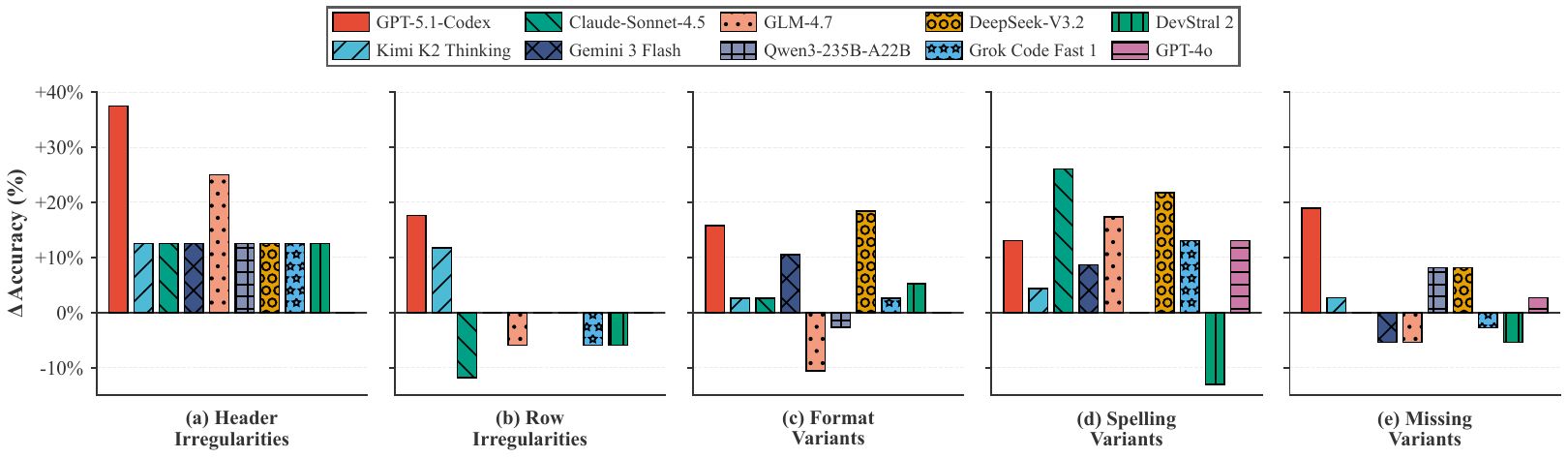}
  \caption{Profiling gains by irregularity type.}
  \Description{Wide grouped chart comparing profiling gains across irregularity categories and models.}
  \label{fig:rq4_dirtytype}
\end{figure*}

\noindent\textbf{Interaction helps, but gains vary by model.}
All agents improved under \interact, with accuracy gains ranging from +4 to +28 points over \orig\ (Figure~\ref{fig:rq3_interact}).
Models that start higher under \orig\ often gain more from interaction.
For example, GPT-5.1-Codex and Kimi K2 Thinking achieved large gains, while GPT-4o improved the least.
DeepSeek-V3.2 is a notable exception, showing a large gain despite its low accuracy under \orig.
Overall, recovery depends not only on code generation quality but also on how well a model identifies ambiguities and resolves them.


\noindent\textbf{More questions do not guarantee better gains.}
A natural hypothesis is that asking more questions leads to higher accuracy, but Figure~\ref{fig:rq3_budget} shows otherwise.
Models varied in how much of the budget they used, yet gains depended on question quality rather than volume.
For example, DevStral 2 consumed over 95\% of the budget for a 20-point gain, while GPT-5.1-Codex gained 25 points using about 60\%.
This suggests that effective agents can prioritize impactful ambiguities and decide when to stop, thereby avoiding unnecessary or redundant questions.

\noindent\textbf{Multi-table alignment ambiguities are hardest to identify.}
Figure~\ref{fig:rq3_bottleneck} shows that multi-table alignment ambiguities have the lowest recall, as models often misinterpret join fields, assuming their understanding aligns with the user's intent.
In contrast, \emph{Group-level concept} has the highest recall, as unclear definitions of concepts such as averages or sums make the ambiguity easier to identify.
DeepSeek-V3.2 performs well across types, suggesting robust identification capabilities.

\begin{takeawaybox}
\textbf{Takeaway III.} Interaction improves accuracy, but question quality matters more than quantity. Models show significant performance differences across ambiguity types, with Multi-table alignment being the hardest to resolve.
\end{takeawaybox}


\subsection{RQ4: Prep-Code Generation}
\label{sec:experiments:rq4}
RQ4 asks whether agents can accurately generate prep-code given an \emph{unambiguous request}.
We compared two settings that both provided $\dreq$ and a small input sample.
In NoProfile, the agent directly performed \emph{Code}.
In \profile, the agent could perform \emph{Profile} on the full input tables before performing \emph{Code}.
This setup isolates irregularity handling and tests whether profiling improves prep-code generation under irregular data.

\noindent\textbf{Data profiling is a double-edged sword.}
Figure~\ref{fig:rq4_overall} shows that data profiling has uneven effects across models.
GPT-5.1-Codex and DeepSeek-V3.2 gained 10 and 4 points, while several models changed little; DevStral~2 and GPT-4o even regressed.
Profiling helps only when summaries are concise, relevant, and correctly used in code.
Otherwise, models may misfire on noisy signals or lose focus, skipping required steps.

\sloppy

\noindent\textbf{Profiling effects vary by irregularity type.}
As shown in Figure~\ref{fig:rq4_dirtytype}, profiling helps stronger models (e.g., GPT-5.1-Codex and Kimi~K2~Thinking) across most irregularity patterns, but the gains vary.
Header irregularities yield the most consistent improvements. 
Without profiling, models often overlook header cleanup even when input samples contain suspicious column names.
\profile directs attention to input validation, and header issues are easy to fix once identified.
In contrast, other irregularity patterns are more diverse and sparse, making them harder to detect and address. 
When exposed, weaker models may misinterpret partial cues and apply harmful or excessive cleaning.
\fussy 

\begin{takeawaybox}
\textbf{Takeaway IV.} Profiling yields uneven gains and its utility varies by model.  
Without explicit instructions, agents often overlook or mishandle data irregularities.
\end{takeawaybox}


\begin{figure}[t]
  \centering
  \makebox[\columnwidth][c]{%
    \includegraphics[width=0.9\columnwidth]{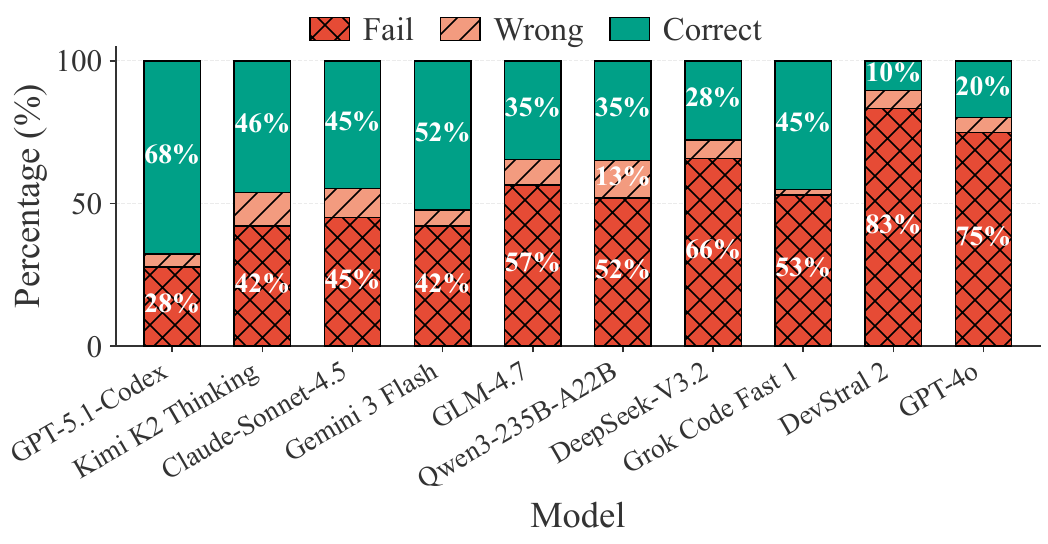}%
  }
  \caption{Translation outcome composition across models.}
  \Description{Stacked chart showing the composition of code-to-workflow translation outcomes for each model.}
  \label{fig:rq5_composition}
\end{figure}

\begin{figure}[t]
  \centering
  \makebox[\columnwidth][c]{%
    \includegraphics[width=0.9\columnwidth]{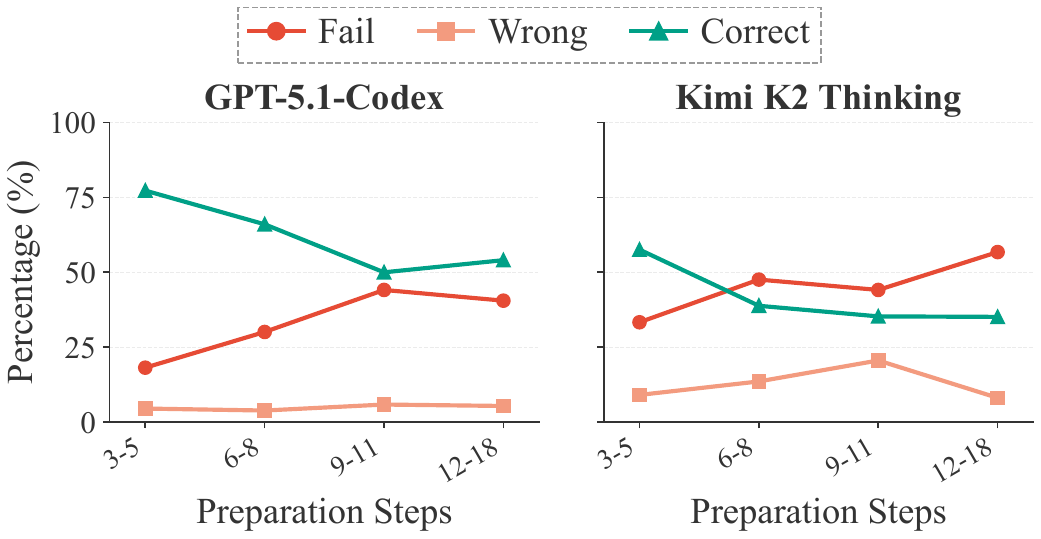}%
  }
  \caption{Translation outcomes by Preparation steps.}
  \Description{Chart showing how translation outcomes vary by the number of preparation steps.}
  \label{fig:rq5_complexity} %
\end{figure}

\subsection{RQ5: Reliability of Workflow Translation}
\label{sec:experiments:rq5}
The preceding experiments focus on code correctness.
In practice, systems also need to generate GUI workflows that users can inspect.
RQ5 evaluates whether models can reliably translate code into executable workflows.
In this setting, the model was given $\gtcode$ together with the workflow operator definitions and could only perform \emph{Translate}.
We executed each translated workflow on $\intab$ and classified its execution outcome as \emph{Fail}, \emph{Wrong}, or \emph{Correct} following \secref{sec:code_to_gui_translation}.


\noindent\textbf{Failures are dominated by non-executable workflows.}
Figure~\ref{fig:rq5_composition} shows that \emph{Wrong} is consistently rare, never exceeding 13\%.
When translation fails, the cause is typically \emph{Fail} rather than an incorrect but executable workflow.
Conversely, workflows that execute are typically \emph{Correct}, indicating that agents grasp the logic but struggle to express it with valid workflow structures.
The primary challenge lies in generating structurally valid workflows.
Since our operators are not seen during pretraining and are learned solely from the prompt, models often produce incorrect bindings or parameters that block execution.

\noindent\textbf{Complexity degrades translation reliability.}
Figure~\ref{fig:rq5_complexity} shows that translation remains imperfect even for short workflows.
At 3–5 steps, GPT-5.1-Codex reached 77.27\% \emph{Correct}, while Kimi~K2~Thinking lagged at 57.58\%.
As workflows grew longer, \emph{Correct} dropped mainly due to rising \emph{Fail} rates, not \emph{Wrong}, indicating that the dominant challenge lies in generating executable workflows.
These failures stem from binding and parameter errors under unfamiliar operator definitions. 
Although Kimi~K2~Thinking is the strongest open-weight model in code-to-workflow translation (Figure~\ref{fig:rq5_composition}), it lags behind GPT-5.1-Codex across all workflow lengths (Figure~\ref{fig:rq5_complexity}).

\begin{takeawaybox}
\textbf{Takeaway V.} Workflow translation remains unreliable: failures are mostly due to non-executable workflows, and increasing complexity further reduces both executability and accuracy.
\end{takeawaybox}

\section{Research Opportunities} \label{sec:research-opportunity}

\noindent\textbf{Improving End-to-End Accuracy.}
End-to-end NL-driven data preparation requires both correct outputs and procedures that users can verify.
RQ1 (\secref{sec:experiments:rq1}) shows that current agents remain unreliable even with clarifications, mainly due to three limitations: (1) detecting ambiguity in user requests, (2) handling irregular inputs, and (3) generating workflows that support user verification.
We outline research opportunities in the following.

\noindent\underline{Interactive Disambiguation}.
A key opportunity is to improve clarification seeking from two angles:
(i) \emph{Foundation-model training} that explicitly rewards asking rather than guessing when requests are underspecified, e.g., via preference optimization or RL for clarification behaviors~\cite{zhang2024modeling}; and
(ii) \emph{Agentic system design} that uses a dedicated disambiguation stage to rewrite a raw request into an explicit specification that records the resolved choices, so downstream execution consumes the specification rather than re-interpreting the original request.

\noindent\underline{Prep-Code Generation}.
NL-to-code pipelines often struggle on irregular tables when they rely only on natural language requests and small samples.
Recent systems leverage profiling and execution feedback to guide iterative refinement~\cite{huang2024cocoon,hong2025data}.
A promising direction is a profile-first approach, where the agent first profiles full tables to extract key properties (e.g., data types, missing values, key constraints), and then uses these properties to guide code generation and add runtime assertions to localize and repair errors~\cite{schelter2018automating}.

\noindent\underline{Code-to-Workflow Translation}.
Workflows are important because users often cannot inspect generated code directly.
One opportunity is an \emph{LLM+compiler} design: the compiler analyzes the generated prep-code to derive an initial workflow, while the LLM fills remaining parameters and generates human-readable annotations. 
The workflow should expose intermediate results to support inspection and iterative edits~\cite{kandel2011wrangler}.

\noindent\textbf{User Profiles for Data Preparation.}
In data preparation, the same user often makes recurring choices across tasks, such as how to handle missing values or resolve duplicates. Asking users to restate these choices repeatedly increases interaction cost, while relying on global defaults can lead to mismatches with user preferences and downstream objectives~\cite{guha2024automated}. This motivates capturing such recurring choices as user profiles and reusing them across tasks.

Designing user profiles for data preparation involves two key aspects.
First, systems need clear rules to determine when to ask for user input, when to store a resolved choice as a profile entry, and when it is appropriate to reuse that entry in a new task.
Second, profile management mechanisms are needed to specify when a stored profile entry should be applied (e.g., under which data schemas, operations, or table characteristics), resolve conflicts across tasks, and update entries as user preferences change over time.

\noindent\textbf{Preparation Beyond Structured Data.} 
\sysname{} focuses on tabular inputs, so its empirical conclusions are limited to tabular data preparation. 
Still, the three capabilities it evaluates suggest natural extensions to semi-structured and unstructured sources such as JSON, logs, and PDFs.
For interactive disambiguation, the multi-table alignment challenge becomes schema-free alignment over entities, fields, or extracted values.
For prep-code generation, profiling would need to summarize nested structures and noisy text, which may provide useful but less reliable signals.
For code-to-workflow translation, workflow systems would need operators beyond relational-style table transformations, including extraction, parsing, normalization, and nested-structure manipulation.
Extending \sysname{} to non-tabular sources is an important direction, and a full treatment is left to future work.

\section{Related Work}
\label{sec:related}

\sloppy

\noindent\textbf{LLM-based Data Preparation.}
Data preparation has long been a core topic in data management research~\cite{kandel2011wrangler,dallachiesa2013nadeef,rekatsinas2017holoclean,rezig2019data}. Recent studies have explored the use of LLMs for data preparation~\cite{chen2025empowering,zhou2026can}, which can be broadly grouped by the role LLMs play in the pipeline.




\noindent\underline{LLM as Data Processor}.
This line of work uses LLMs directly as data processing operators.
LLM-GDO~\cite{ma2023llms} applies LLMs to single-step data transformations, and other work embeds them as preprocessing operators in data pipelines~\cite{zhang2023large}.
SEED~\cite{chen2023seed} performs semantic annotation and data curation, COMEM~\cite{wang2025match} performs record-level matching, and CHORUS~\cite{kayali2023chorus} generates metadata for table discovery.
Our work instead studies LLMs as an NL interface.

\noindent\underline{LLM as NL Interface}.
This line of work uses LLMs to translate NL requests into executable data preparation actions.
SheetCopilot~\cite{li2023sheetcopilot}, SheetAgent~\cite{chen2025sheetagent}, and Dango~\cite{chen2025dango} operate within spreadsheet or GUI environments, mapping user intents to application-specific actions.
CleanAgent~\cite{qi2024cleanagent} targets data standardization with tool invocation and iterative refinement.
AutoPrep~\cite{fan2024autoprep} and HAIPipe~\cite{chen2023haipipe} generate task-specific pipelines for downstream question answering or machine learning.
LLMs have been applied to data integration, including semantic matching~\cite{peeters2023entity}, schema matching~\cite{parciak2024schema}, and harmonization pipelines~\cite{qiang2023agent,santos2025interactive}.
A central challenge in NL interfaces is handling ambiguous user requests. 
In Text-to-SQL, \emph{Schema Linking} and \emph{Value Linking} identify which columns and values a query refers to, and prior work has also studied data ambiguity in input tables~\cite{huang2023data,veltri2023data}.
These problems correspond to the \emph{Data Interpretation} category in our taxonomy.
However, NL-driven data preparation introduces additional ambiguity types because users are not only querying an existing schema, but also defining multi-step transformations.
\sysname{} therefore also covers \emph{Concept Interpretation} and \emph{Operational Interpretation}.
Together, these two categories account for 83.2\% of ambiguities in \sysname{}, and our experiments show that they substantially affect prep-code accuracy.
Unlike prior work that builds agents for specific data preparation tasks, \sysname{} targets NL-driven data preparation in general and introduces a benchmark that systematically evaluates its core capabilities.

\noindent\textbf{LLM4DATA Benchmarks.}
Benchmarks play a central role in understanding the capabilities and limitations of LLMs for data-centric tasks. Recent surveys provide comprehensive overviews of benchmarks for evaluating LLMs on data-centric tasks over structured, semi-structured, and unstructured inputs~\cite{zhou2025survey,zhu2025survey}.
Building on these surveys, we group representative benchmarks according to the primary task they evaluate. For \emph{querying and reasoning}, widely used benchmarks include Spider~\cite{yu2018spider}, BIRD~\cite{li2023can}, and TabFact~\cite{chen2019tabfact}. Beyond querying, recent benchmarks increasingly adopt \emph{execution-based} evaluation and define tasks over concrete \emph{artifacts}, such as spreadsheets, code, and data pipelines.
In spreadsheet-centric settings, TEMPTABQA~\cite{gupta2023temptabqa} studies table reasoning with temporal constraints, and SpreadsheetBench~\cite{ma2024spreadsheetbench} evaluates realistic spreadsheet manipulation tasks. For data analysis, DS-1000~\cite{lai2023ds} and DA-Code~\cite{huang2024code} benchmark executable data-science coding, while DA-Bench~\cite{hu2024infiagent}, DABstep~\cite{egg2025dabstep}, and DSEval~\cite{zhang2024benchmarking} focus on multi-step analysis with execution-based verification.
At the pipeline level, Spider~2.0~\cite{lei2024spider} moves Text-to-SQL toward more realistic, workflow-oriented settings, ELT-Bench~\cite{jin2025elt} benchmarks end-to-end ELT construction, Harmonia~\cite{santos2025interactive} evaluates mixed-initiative data harmonization workflows, and KramaBench~\cite{lai2025kramabench} benchmarks end-to-end data pipelines built from real artifacts.
In comparison, existing benchmarks either target different problem settings or evaluate individual capabilities in isolation. PrepBench is designed to systematically evaluate all three core capabilities required for NL-driven data preparation: interactive disambiguation, prep-code generation, and code-to-workflow translation.

\fussy

\section{Conclusion}
\label{sec:conclusion}

This paper examined NL-driven data preparation and evaluated how far current LLM-based agents are from supporting it in realistic settings. We presented \sysname, a benchmark systematically constructed from real-world Preppin\textquotesingle{} Data challenges. We converted each challenge into a benchmark task with \gtcode, \dreq, \dkb and \gtflow.  We designed three execution modes that enabled both end-to-end evaluation and targeted analysis of individual capabilities. We analyzed various aspects of \sysname, including task complexity, ambiguity types, data irregularities, and operator usage. These statistics characterized the coverage and difficulty of the benchmark.

We evaluated ten state-of-the-art proprietary and open-weight LLMs on \sysname. The results showed that NL-driven data preparation remained challenging for current models. Firstly, the best-performing model (GPT-5.1-Codex) achieved 54.9\% accuracy on end-to-end prep-code generation. Removing ambiguity increased accuracy to 85.3\%, suggesting that ambiguous user requests were a major source of error. Secondly, interactive disambiguation improved accuracy, but its effectiveness was limited by the quality of clarification questions, which were often incomplete or ineffective. Thirdly, code-to-workflow translation was also challenging, as it required the model to reason over explicit operator definitions provided in the prompt, rather than relying solely on prior knowledge learned during pretraining. Fourthly, model cost did not consistently correlate with performance; for prep-code generation, Gemini~3~Flash nearly matched the best accuracy at less than one-fifth of the cost. 
Overall, our study quantitatively characterizes the limitations of current LLM-based agents and identifies the capability gaps that need to be addressed to support NL-driven data preparation in practice.
We released \sysname, the evaluation framework, and baseline implementations at
\url{https://github.com/TsinghuaDatabaseGroup/prepbench}.


\bibliographystyle{ACM-Reference-Format}
\bibliography{refs}

\end{document}